%% file: sample-acmsmall-conf.tex
\documentclass[sigconf,screen]{acmart}

\AtBeginDocument{%
  \providecommand\BibTeX{{%
    \normalfont B\kern-0.5em{\scshape i\kern-0.25em b}\kern-0.8em\TeX}}}

\usepackage{amsmath,amsfonts}
\usepackage{microtype}
\usepackage{multirow}
\usepackage{caption}
\usepackage{multirow}
\usepackage{textcomp}
\usepackage{xcolor}
\usepackage{framed}
\usepackage{listings}
\usepackage{graphicx,subfigure}
\usepackage{bbding}
\usepackage{color}
\usepackage{makecell}
\usepackage{enumitem}
\usepackage{xspace}
\usepackage{tcolorbox}
\usepackage{threeparttable}
\usepackage{verbatim}
\usepackage{hyperref}
\usepackage{courier} 
\usepackage{listings, xcolor}
\usepackage[linesnumbered,ruled,vlined]{algorithm2e}
\usepackage{lipsum} 
\usepackage{booktabs} 
\usepackage{listings}
\usepackage{xcolor}  

\newcommand{\tool}{\textit{FT2Ra}\xspace}

\newcommand{\qi}[1]{{\textcolor{black}{#1}}}

\setcopyright{acmlicensed}
\acmDOI{10.1145/3650212.3652130}
\acmYear{2024}
\copyrightyear{2024}
\acmSubmissionID{issta24main-p700-p}
\acmISBN{979-8-4007-0612-7/24/09}
\acmConference[ISSTA '24]{Proceedings of the 33rd ACM SIGSOFT International Symposium on Software Testing and Analysis}{September 16--20, 2024}{Vienna, Austria}
\acmBooktitle{Proceedings of the 33rd ACM SIGSOFT International Symposium on Software Testing and Analysis (ISSTA '24), September 16--20, 2024, Vienna, Austria}
\received{16-DEC-2023}
\received[accepted]{2024-03-02}

\begin{document}
\title{FT2Ra: A Fine-Tuning-Inspired Approach to Retrieval-Augmented Code Completion}

\author{Qi Guo}
\orcid{0009-0008-8002-8068}
\authornote{This work was done during the author's visit to Singapore Management University.}
\affiliation{%
  \institution{Tianjin University}
  \city{Tianjin}
  \country{China}
}
\email{bxguoqi@tju.edu.cn}

\author{Xiaohong Li}
\orcid{0000-0002-0752-6764}
\affiliation{%
  \institution{Tianjin University}
  \city{Tianjin}
  \country{China}
}
\email{xiaohongli@tju.edu.cn}

\author{Xiaofei Xie}
\orcid{0000-0002-1288-6502}
\affiliation{%
  \institution{Singapore Management University}
  \country{Singapore}
}
\email{xfxie@smu.edu.sg}

\author{Shangqing Liu}
\orcid{0000-0002-5598-4006}
\authornote{Corresponding author}
\affiliation{%
  \institution{Nanyang Technological University}
  \country{Singapore}
}
\email{liu.shangqing@ntu.edu.sg}

\author{Ze Tang}
\orcid{0009-0002-8062-9986}
\affiliation{%
  \institution{Nanjing University}
  \city{Nanjing}
  \country{China}
}
\email{zetang@smail.nju.edu.cn}

\author{Ruitao Feng}
\orcid{0000-0001-9080-6865}
\affiliation{%
  \institution{Singapore Management University}
  \country{Singapore}
}
\email{rtfeng@smu.edu.sg}

\author{Junjie Wang}
\orcid{0009-0002-3847-6760}
\affiliation{%
  \institution{Tianjin University}
  \city{Tianjin}
  \country{China}
}
\email{junjie.wang@tju.edu.cn}

\author{Jidong Ge}
\orcid{0000-0003-1773-0942}
\affiliation{%
  \institution{Nanjing University}
  \city{Nanjing}
  \country{China}
}
\email{gjd@nju.edu.cn}

\author{Lei Bu}
\orcid{0000-0003-0517-7801}
\affiliation{%
  \institution{Nanjing University}
  \city{Nanjing}
  \country{China}
}
\email{bulei@nju.edu.cn}


\begin{abstract}
The rise of code pre-trained models has significantly enhanced various coding tasks, such as code completion, and tools like GitHub Copilot. However, the substantial size of these models, especially large models, poses a significant challenge when it comes to fine-tuning them for specific downstream tasks. As an alternative approach, retrieval-based methods have emerged as a promising solution, augmenting model predictions without the need for fine-tuning.
Despite their potential, a significant challenge is that the designs of these methods often rely on heuristics, leaving critical questions about \textit{what information should be stored or retrieved and how to interpolate such information for augmenting predictions}.

To tackle this challenge, we first perform a theoretical analysis of the fine-tuning process, highlighting the importance of $\Delta logits$ as a catalyst for improving model predictions. Building on this insight, we develop a novel retrieval-based method, \textit{FT2Ra}, which aims to mimic genuine fine-tuning. While \textit{FT2Ra} adopts a retrieval-based mechanism, it uniquely adopts a paradigm with a learning rate and multi-epoch retrievals, which is similar to fine-tuning.

We conducted a comprehensive evaluation of \textit{FT2Ra} in both token-level and line-level code completions. Our findings demonstrate the remarkable effectiveness of \textit{FT2Ra} when compared to state-of-the-art methods and its potential to genuine fine-tuning.
In token-level completion, which represents a relatively easier task, \textit{FT2Ra} achieves a 4.29\% improvement in accuracy compared to the best baseline method on UniXcoder. In the more challenging line-level completion task, we observe a substantial $\sim$2$\times+$ increase in Exact Match (EM) performance, indicating the significant advantages of our theoretical analysis. Notably, even when operating without actual fine-tuning, \textit{FT2Ra} exhibits competitive performance compared to the models with real fine-tuning.
\end{abstract}




\maketitle


\section{Introduction}

\input{sections/introduction}

\section{Background and Problem}

\input{sections/background}
\subsection{Problem}
\input{sections/motivation}

\section{Approach}
\input{sections/approach}

\section{Experimental Setup}
\input{sections/setup}

\section{Experimental Results}

\input{sections/results}

\section{Threats to Validity}
\input{sections/discussion}

\section{Related Work}
\input{sections/related}

\section{Conclusion}
\input{sections/conclusion}
\section{DATA AVAILABILITY}
Our source code and experimental data are available at~\cite{ft2rawebsite}.
\section{Acknowledgment}
This work was partially supported by the National Key R\&D Project (2021YFF1201102), the National Natural Science Foundation of China (Key Program, Grant No. 62332005), the National Key R\&D Program of China (2021ZD 0112903), the National Natural Science Foundation of China (Grant No. 61872262 and No. 62102283), the National Research Foundation, Singapore, and the Cyber Security Agency under its National Cybersecurity R\&D Programme (NCRP25-P04-TAICeN). Lei Bu is supported in part by the Leading-edge Technology Program of Jiangsu Natural Science Foundation (No. BK20202001), the National Natural Science Foundation of China (No. 62232008, 62172200). Any opinions, findings and conclusions or recommendations expressed in this material are those of the author(s) and do not reflect the views of National Research Foundation, Singapore and Cyber Security Agency of Singapore.
\bibliographystyle{ACM-Reference-Format}
\bibliography{sections/reference}

\end{document}

%% file: sections/introduction.tex
In the realm of software engineering, code pre-trained models (CPMs) specialized for code generation and completion are becoming increasingly prevalent.
Recently, a series of code completion plugins such as GitHub Copilot~\cite{githubcopilot}, and Visual Studio IntelliCode~\cite{intellicode} have significantly alleviated the burden on software developers and enhanced their development efficiency.  

Code-centric pre-trained models are generally trained using vast amounts of source code data harvested from open repositories. In the inference phase, CPMs typically map the code prefixes to fixed-sized representations and use the representations to predict the next code token.
However, despite the extensive training data, CPMs still struggle to capture rare or specialized patterns. On one hand, the rarity of certain patterns in the training data makes it difficult for the model to learn them adequately. On the other hand, the complex inter-dependencies between different data samples could include conflicting coding styles or logic that the model fails to reconcile. Furthermore, these CPMs may not excel in specific domains where task-oriented or project-specific knowledge is essential, such as project-specific API invocations. For example, the recent study~\cite{zhang2023repocoder} disclosed that these general-purpose pre-trained models are inferior in repository-level code completion where the interrelated dependencies among files within a repository are missed for these general models. 
Therefore, the post-training enhancement of these models becomes a crucial task.

To tackle the outlined challenges, a straightforward approach is to fine-tune the pre-trained models using specialized data, such as missing patterns or project-specific information. However, fine-tuning comes with its own set of limitations, particularly concerning the computational resources required and the quality of data necessary for effective adjustment. Fine-tuning the entire model necessitates storing and updating a colossal parameter set, an operation that becomes increasingly costly and often infeasible as model size escalates into billions of parameters. Furthermore, the success of this strategy hinges on the availability of high-quality, task-specific data. 
While parameter-efficient fine-tuning techniques have been proposed~\cite{hu2021lora, liu2023gpt, lester2021power}, they still demand considerable computational resources for fine-tuning.

Recent research~\cite{zhang2023repocoder, Khandelwal2020Generalization, tang2023domain} proposes an alternative route through the use of retrieval-augmented language models (RaLMs). These models supplement the capabilities of pre-trained models by incorporating retrieval mechanisms that source information (e.g., rare patterns) from an external database, thereby bypassing the need for additional fine-tuning. This method enables the model to explicitly store and retrieve rare patterns, as opposed to implicitly integrating them into the model's parameters~\cite{Khandelwal2020Generalization}. This paradigm aligns well with human learning behavior, where sparse examples are leveraged to generalize effectively to new situations. Empirical results demonstrate that RaLMs can significantly enhance the performance of CPMs, particularly in the prediction of rare patterns.

For retrieval-augmented language models, two main challenges exist the identification of similar samples from an external database and the effective utilization of this retrieved information for making predictions. Typically, the former is addressed by retrieving nearest neighbors based on distance metrics in a pre-trained embedding space. For the latter, existing methods adopt different methods such as employing frequency analysis~\cite{WikipediaFrequency} and empirical probabilities~\cite{WikipediaEmpiricalProb} to integrate the retrieved information. For example, kNN-LM~\cite{Khandelwal2020Generalization} retrieves the neighbors and computes a distribution
over neighbors based on a softmax of their negative distances, which are used to augment the original predictions. The most recent work 
kNM-LM~\cite{tang2023domain} retrieves the code tokens that the language model fails to predict and normalize into a distribution, which is merged with the predictions of the language model. While these heuristic approaches have yielded promising results, they largely depend on empirical settings, leaving theoretical gaps in terms of what information should be retrieved and how this information can be better exploited.

In this paper, to better understand the optimal use of retrieved knowledge, we first conduct a theoretical analysis of the fine-tuning process in CPMs. Our theoretical analysis and derivation reveal insights for designing a strategy that more closely approximates the effects of fine-tuning. While our theoretical derivation does incorporate certain approximations, the evaluation results still demonstrate the effectiveness of the retrieval mechanism. 
Specifically, our analysis indicates that the logits discrepancy between the predicted and actual values associated with neighbors (i.e., $\Delta logits$)~\footnote{\qi{$\Delta logits$ represents the difference in logits before and after gradient descent, which can be expressed as $\Delta logits = logits' - logits$.}} serves as crucial information for augmenting CPM predictions. Based on the analysis, we develop a novel retrieval-augmentation technique, denoted as \tool, for code completion tasks. CPMs can recalibrate and improve its predictions by adding the $\Delta logits$ to its logit output. 
Furthermore, akin to the iterative nature of the fine-tuning process, \tool is designed to operate through an iterative retrieval cycle, progressively updating the external database to refine the quality of retrieved information, thereby continuously improving prediction accuracy.

To showcase the effectiveness of \tool, we selected four state-of-the-art retrieval-based methods: kNN-LM~\cite{Khandelwal2020Generalization}, kNM-LM~\cite{tang2023domain}, ReACC~\cite{lu2022reacc}, and BM25~\cite{robertson2009probabilistic}. Our evaluation encompassed both token-level and line-level code completion. The experimental results demonstrate that, guided by our theoretical analysis, \tool significantly outperforms the baseline methods and achieves competitive performance similar to actual fine-tuned models.
For instance, in the context of token-level completion, \tool obtains an average accuracy of 74.22\% (\textbf{4.29\%+}) on UniXcoder, whereas UniXcoder and the top-performing baseline, kNM-LM, achieve accuracy of 54.07\% and 69.93\%, respectively. In the more challenging line-level completion task, \tool achieves an average Exact Match (EM) score of 26.32 ($\sim$2$\times+$ ) on UniXcoder. In contrast, UniXcoder and kNM-LM only manage scores of 1.63 and 13.93, respectively.
We also observed that, in line-level completion using UniXcoder, \tool achieves performance better than that of the fine-tuned UniXcoder model after 10 epochs, even when operating without fine-tuning. 
These results not only demonstrate the effectiveness of \tool but also highlight its significant potential to achieve competitive results comparable to those of fine-tuned models. Furthermore, our additional evaluations reveal that the iterative retrieval mechanism designed within \tool significantly contributes to its performance.

In summary, this paper makes the following contributions:

\begin{itemize}[leftmargin=*]
\item 
\textbf{Theoretical Analysis:} We perform a theoretical analysis of the model fine-tuning process. This analysis provides valuable insights into \textit{how to effectively exploit retrieval information in retrieval augmentation mechanisms}.

\item 
\textbf{Methodology:} Building upon the insights derived from our theoretical analysis, we introduce a novel method called \tool. This innovative approach emulates real fine-tuning through an iterative retrieval process, enhancing its effectiveness.
\item
\textbf{Comprehensive Evaluation:}
We conduct an extensive evaluation to evaluate the effectiveness of \tool in both token-level and line-level code completion tasks. The results highlight substantial improvements achieved by \tool.
\item
\textbf{Open-Source Resources:}
We have made the pertinent data, detailed experimental findings, and the tools publicly available~\cite{ft2rawebsite}.

\end{itemize}


%% file: sections/background.tex

\subsection{Retrieval-Augmented Language Models}\label{sec:background-retrieval}
Recently, a series of retrieval-augmented language models~\cite{tang2023domain, Khandelwal2020Generalization, guu2020retrieval} have been proposed to augment language models with external knowledge~\cite{yang2023leandojo,chen2022decoupling,hofstatter2023fid}. 
Retrieval-augmented techniques can generally be divided into two types.
The first type is at the input layer~\cite{ram2023context,izacard2022few,guu2020retrieval}, where the retrieved information is text chunks.
The second type is at the output layer~\cite{Khandelwal2020Generalization,tang2023domain,alon2022neuro}, where the retrieved information is tokens. By combining the retrieved tokens with the tokens generated by the original model, the accuracy of the retrieval-augmented model's generation for each token can be improved.
The first type of method can provide the model with more external knowledge, making it adept at handling tasks in the NLP field such as knowledge-based question answering~\cite{lewis2020retrieval,wang2023learning,shi2023replug}. The second type of method can refer to the retrieved information to correct the generated tokens, making it more suited for handling strictly structured generative tasks, such as code completion~\cite{drozdov2022you,alon2022neuro,de2021mention}.
In this work, we mainly focus on the second category. 

To better understand the mechanism, we take kNN-LM~\cite{Khandelwal2020Generalization} as an example for a detailed explanation. Given a context sequence $c_t = (w_1, \dots, w_{t-1})$, the language models (LMs) estimate $p_{LM}(w_t|c_t)$, i.e., the probability distribution over the next token $w_t$. kNN-LM is designed to augment a pre-trained language model with a set of nearest neighbours retrieved from an external text collection, which can be the training set $D$. Different from fine-tuning, retrieval augmentation does not need any retraining.
In particular, RaLM includes two tasks, i.e., building a datastore and retrieval-augmented inference.

\noindent\textbf{Datastore}: The datastore is a retrieval set, which can be built with a forward pass by LM on the prepared text collection to store the context-target pairs as the subject of a query. We denote a function $f(\cdot)$ to map a context $c$ to a fixed-length vector representation computed by a pre-trained LM. Given an example $(c_i, w_i) \in D$, we can pass $c_i$ to a LM to get its vector representation, i.e., $k_i=f(c_i)$. The dataset $D$ is a set of datasets such as the training data or other domain-specific data. In this way, we can obtain the key-value pair $(k_i, v_i)$, where $k_i$ is the context representation computed from LM and $v_i$ is the target word $w_i$. Hence, the datastore $(K, V)$ is a set of all context-target pairs built from $D$, which can be expressed as:
\begin{equation}
(K, V) = \{(f(c_i), w_i) \mid (c_i, w_i) \in {D}\}
\end{equation}
\textbf{Inference}: The inference phase includes neighbour retrieval and the use of neighbour prediction information.
Given a new input $x$, the model first computes its context representation i.e., $f(x)$. Using $f(x)$ as a query to retrieve the $k$-nearest neighbours $\mathcal{N}$ from the datastore $(K, V)$ based on a defined distance function $dis(\cdot)$ such as Euclidean distance. Then it computes a distribution over these $k$ neighbors using a softmax function. The probability for each vocabulary item is aggregated across all occurrences in the retrieved targets. Note that the items in the vocabulary set that do not appear in the retrieved targets have a probability of zero.
\begin{equation}
\label{eq-retrieval}
p_{kNN}(y|x) \propto \sum_{(k_i,v_i) \in \mathcal{N}} \mathbf{1}_{\{y = v_i\}} \exp(-dis(k_i, f(x)))
\end{equation}
The final distribution is interpolated with the original LM distribution $p_{LM}(y|x)$ and $p_{kNN}(y|x)$ to obtain the joint distribution:
\begin{equation}
\label{eq-knn}
p(y|x) = (1-\lambda)p_{LM}(y|x) + \lambda p_{kNN}(y|x)
\end{equation}
where $\lambda$ is a tuned hyper-parameter to control the weight of generation and retrieval.


%% file: sections/motivation.tex

From Equation~\ref{eq-knn}, we can observe that the final distribution of kNN is the weighted sum of the original LM distribution i.e., $p_{LM}(y|x)$ and the retrieved nearest neighbor distribution i.e., $p_{kNN}(y|x)$.  The key problem is how to interpolate the retrieved knowledge in the prediction, i.e., the design of $p_{kNN}(y|x)$. In kNN-LM, $p_{kNN}(y|x)$ is computed from Equation.~\ref{eq-retrieval} based on negative distances and the aggregated probability for each vocabulary item across all its occurrences in the retrieved targets. While the design is intuitive,  it is still based on heuristics and lacks theoretical analysis and explanation. A key question to \textit{identify what kinds of information should be retrieved and how best to leverage that information.}

%% file: sections/approach.tex
In this section, we delve into a theoretical analysis aimed at identifying useful retrieval information, drawing inspiration from the fine-tuning process commonly employed for enhancing the performance of CPMs. Subsequently, we introduce our method, \tool, which focuses on the effective interpolation of this retrieved information to improve the predictive accuracy of CPMs.

\subsection{Inspiration From Fine-tuning}\label{sec:inspiration}

Fine-tuning serves as a practical technique for boosting the performance of pre-trained models, particularly when applied to domain-specific tasks or datasets that the original model may not adequately cover. Our goal is to distil insights from the mechanics of fine-tuning to inform the design of a retrieval-augmented method that approximates the performance improvements seen with fine-tuning, yet obviates the need for the fine-tuning process itself.

Let $\mathcal{M}$ represent a given language model capable of predicting the next token  $x_t$ based on its preceding context sequence $x = (x_1, x_2, \ldots, x_{t-1})$. We proceed with the following definitions:

\begin{itemize}[leftmargin=*]
  \item $\theta$ denotes the trained model parameters of   $\mathcal{M}$.
    \item $y$ is the ground-truth for $x_t$ as a one-hot encoding, where the index corresponding to $x_t$
  is marked as 1 while other indices are 0.  $y \in \mathbb{R}^{v}$ is a vector where $v$ is the length of the vocabulary set. 
    \item $y'$ is the model prediction result for the next token, i.e., $y'=\mathcal{M}(x|\theta)$ and $y' \in \mathbb{R}^{v}$, which denotes the predicted probability of each token in the vocabulary set with the context. Typically, $y'$ is the output for the probability layer of the model $\mathcal{M}$. 
    \item $logits  \in \mathbb{R}^{v} $ encapsulates the values in the logits layer, preceding the probability layer.
    \item $seqout \in \mathbb{R}^{dmodel} $ is the output of the decoder sequence output layer, preceding the logits layer, and $dmodel$ is the dimension of this layer.

\end{itemize}

Suppose the LM $\mathcal{M}$  undergoes fine-tuning through multiple epochs, following best practices. Without loss of generality, we assume that the loss for a given input $x$ diminishes after each iteration of the fine-tuning (i.e., the gradient descent algorithm). Let $\theta$ and $\theta'$ denote the model's parameters before and after an epoch of fine-tuning, respectively, such that $\theta' = \theta + \Delta\theta$. The corresponding loss values before and after the fine-tuning are:
$$
l=\mathcal{L}(\mathcal{M}(x|\theta), y),
l'=\mathcal{L}(\mathcal{M}(x|\theta'), y)
$$

where $\mathcal{L}$ is the loss function, and $y$ is the ground truth for the given context sequence $x$. We define the change in the loss as  $\Delta l = l'-l$. 
Given that the language model is differentiable, the change in loss $\Delta l $ can be expressed as:
\begin{equation}
\label{eq:deta-loss1}
    \begin{aligned}
    \Delta l &= \mathcal{L}(\mathcal{M}(x|\theta + \Delta\theta), y) - \mathcal{L}(\mathcal{M}(x|\theta), y)
    \end{aligned}
\end{equation}

In gradient descent, the learning rate $\eta_{\theta}$
 controls the magnitude of parameter updates:
\begin{equation}
\label{eq:deta-parameter}
\Delta \theta = - \eta_{\theta} \times \frac{\partial \mathcal{L}}{\partial \theta}
\end{equation}

On the other hand, the loss can also be formulated in terms of logits, $l=\mathcal{L}(softmax(logits), y)$, where $logits$ is the model output on $x$ and $y$ is the ground truth. After one iteration of the gradient descent, the loss value $l'$ can be described as $l'=\mathcal{L}(softmax(logits'), y)$, with $logits'$ denoting the output of the updated model. 
Let  $\Delta logits = logits' - logits$, and we can derive:

\begin{equation}
    \label{eq:deta-loss4}
    \begin{aligned}
    \Delta l &= \mathcal{L}(softmax(logits + \Delta logits), y) - \mathcal{L}(softmax(logits), y) \\
             &= (\Delta logits)^T \cdot \frac{\partial \mathcal{L}}{\partial logits}
    \end{aligned}
\end{equation}

Intuitively, if we can develop a method for approximating  $\Delta logits$ without actually engaging in fine-tuning, then these approximated $\Delta logits$ could be directly interpolated into the predictions of the model. This mimics the effects of fine-tuning and may achieve comparable performance, depending on the accuracy of the $\Delta logits$ approximation. 

We observe the final LM-head layer of the generative model, where $logits = lm\_head(seqout)$. We ignore the activation layer in LM-head and can approximately treat the LM-head as a linear layer, from which we can derive:
\begin{equation}
\label{eq:logits}
logits \approx W \cdot seqout + b
\end{equation}
where $W$ is a weight matrix with the dimension $v$ * $dmodel$.

From equation~\ref{eq:logits}, using the chain rule for differentiation~\cite{stanfordcs229}, we get:
\begin{equation}
\label{eq:chain-rule}
\frac{\partial l}{\partial W} = \frac{\partial l}{\partial logits} \cdot seqout^T
\end{equation}

During the gradient descent process, since \(W\) is a part of \(\theta\), according to equation~\ref{eq:deta-parameter}, it also follows the gradient descent rule:
\begin{equation}
\label{eq:gradient-w}
\Delta W = -\eta_{\theta} \cdot \frac{\partial l}{\partial W}
\end{equation}

When we fix the parameters preceding $seqout$ and only fine-tune the subsequent parameters of the model, then according to equation~\ref{eq:logits},~\ref{eq:chain-rule} and~\ref{eq:gradient-w}, we make an approximation:

\begin{equation}
    \label{eq:approx}
\begin{aligned}
\Delta \text{logits} &\approx \Delta W \cdot seqout \\
                    &= -\eta_{\theta} \cdot \frac{\partial l}{\partial W} \cdot seqout \\
                    &= -\eta_{\theta} \cdot \frac{\partial l}{\partial logits} \cdot seqout^T \cdot seqout \\
                    &= -\eta_{\theta} \cdot ||seqout||^2_2 \cdot \frac{\partial l}{\partial logits} 
\end{aligned}
\end{equation}
We use $||seqout||_2$ to denote the L2 norm of $seqout$. Furthermore, we define $-\eta_{logits}$ as  $-\eta_{\theta} \cdot ||seqout||^2_2$, and we can get:
\begin{equation}
\label{eq:logitsresult}
\Delta logits \approx - \eta_{logits} \times \frac{\partial \mathcal{L}}{\partial logits} 
\end{equation}

Equation~\ref{eq:logitsresult} offers a feasible methodology for calculating changes in logits, which can be employed to bolster the current model's performance on $x$ reducing its loss.  To obtain the value of $\frac{\partial \mathcal{L}}{\partial logits}$, we propose the retrieval-based method detailed in Section~\ref{sec:approxlogits}. 

Our derivation implies new insights about what kind of information should be stored and retrieved (i.e., the $\Delta logits$) and how to leverage the information (i.e., add $\Delta logits$ to the predicted logits). In summary, it introduces the following benefits: 1) the retrieval mechanism is theoretically grounded, different from the existing mere heuristic approaches, 2) the retrieval mechanism tries to mimic the fine-tuning process, which has a high potential to achieve high performance and 3) the retrieved knowledge regarding $\Delta logits$ is more fine-grained compared to existing methods, and its integration into the prediction process is both straightforward and direct.

\subsection{Algorithm}\label{sec:approach}
Building on the theoretical insights from fine-tuning, we introduce a novel retrieval-augmented method.

\subsubsection{Approximation of $\frac{\partial \mathcal{L}}{\partial logits}$}
\label{sec:approxlogits}
To approximate the value of $\frac{\partial \mathcal{L}}{\partial logits}$ shown in Equation~\ref{eq:logitsresult}, 
we employ the nearest $k$ neighbors of the sample $x$ for the estimation. The approximation is formulated as
\begin{equation}
\label{eq:approx}
\frac{\partial \mathcal{L}}{\partial logits} \approx \sum_i \lambda_i \times \frac{\partial \mathcal{L}_i}{\partial logits_i}
\end{equation}
where $ 1 \le i \le k$ represents the $i^{th}$ neighbor and $\lambda_i$ serves as a hyper-parameter to adjust the contribution of each neighbor to the approximation. 
Since $\frac{\partial \mathcal{L}_i}{\partial logits_i}$ is the partial derivative with respect to the logits layer, we have $\frac{\partial \mathcal{L}_i}{\partial logits_i} = y'_i-y_i$ for each neighbor.
Incorporating this into Equation~\ref{eq:approx} yields:
\begin{equation}
\label{eq:approx2}
\frac{\partial \mathcal{L}}{\partial logits} \approx \sum_i \lambda_i \times ( y'_i - y_i) 
\end{equation}

Finally, we integrate Equation~\ref{eq:approx2} into Equation~\ref{eq:logitsresult} to derive:
\begin{equation*}
\qi{\Delta logits \approx - \eta_{logits} \times \sum_i \lambda_i \times ( y'_i - y_i) }
\end{equation*}
\begin{equation}
\label{eq:finally}
logits' \approx logits - \eta_{logits} \times \sum_i \lambda_i \times ( y'_i - y_i) 
\end{equation}

Given an input 
$x$, Equation~\ref{eq:finally} offers a mechanism to calculate new logits by leveraging both the original prediction and the contributions from the nearest neighbours.

\subsubsection{Datastore Construction}\label{sec:datastore}
As with prior work in this area~\cite{tang2023domain, Khandelwal2020Generalization}, a retrieval set, referred to as datastore $D$, is essential for storing context-target pairs, often represented as key-value pairs. The nature of the knowledge encapsulated in this datastore depends on the specific retrieval mechanism employed, particularly the type of information used for the calculation.

In the datastore, each key is generated to facilitate distance calculation between the given input and the elements in the retrieval set. For a given training example $(c_i, w_i) \in D$, we map the context $c_i$ to a fixed-length vector representation using a function $f(\cdot)$. In line with previous research~\cite{Khandelwal2020Generalization}, we utilize the last hidden states (i.e., the output of the final layer of the CPM as this mapping function $f(\cdot)$. Hence, the \textit{key} for each entry is $k_i = f(c_i)$.

Considering Equation~\ref{eq:finally}, the \textit{value} associated with each key should include both the ground truth $y_i$ (which corresponds to $w_i$ in a one-hot encoded format) and the predicted probability distribution $y_i'$. Instead of storing the probability vector, we opt to store the corresponding logits vector $logits_i$. This is because: 1) The predicted probability $y_i'$ can be easily recalculated from $logits_i$ whenever needed and 2)  Storing $logits_i$  allows for their use in multiple retrieval iterations, as will be detailed in Section~\ref{sec:mainalgo}.
Given these considerations, the datastore is formally defined as:
\begin{equation*}
(K, V) = \{(f(c_i), (y_i, logits_i)) \mid (c_i, w_i) \in {D}\}
\end{equation*}

\subsubsection{Iterative Nearest Neighbor Retrieving}\label{sec:mainalgo}
\begin{algorithm}[!t]
\DontPrintSemicolon
\KwIn{Test sample: $x$, large model: $\mathcal{M}$, learning rate: $\eta_{logits}$, datastore: $(K,V)$, \\ the number of neighbours: $N$, the number of iterations: $E$}
\KwOut{The output: $y'_x$}
$r\leftarrow f_\mathcal{M}(x)$\; \label{algo:getkey}
$(Y, L, D)\leftarrow retrieve(r, N, (K,V))$\; \label{algo:getpair}
\For{$e \in \{1, \ldots, E\}$}{\label{algo:startepoch}
 $\Delta {logits} \leftarrow 0$\; 
 $logits_x \leftarrow cal\_logits(\mathcal{M}, x)$\; \label{algo:getlogits}
  \For{$(y_i, logits_i) \in (Y, L)$}{\label{alg:nblogitstart}
        $y_i' \leftarrow softmax(logits_i)$\;
        $\lambda_i \leftarrow cal\_weight(d_i, D)$ \; \label{algo:diffweights}
        $\Delta{logits}_i \leftarrow \lambda_i \times \eta_{logits}(\text{y}_i - \text{y'}_i)$\;
        $\Delta{logits} \leftarrow \Delta{logits} + \Delta{logits}_i$\;
    }\label{alg:nblogitend}
    $logits_x \leftarrow logits_x + \Delta{logits} $\; \label{algo:finallogit}
    \For{$l_i \in L$}{
    $l_i \leftarrow l_i + \Delta{logits} $\; \label{algo:updatestore}
    }
}\label{algo:endepoch}
$y'_x = {softmax}({logits}_x)$\;
\textbf{return} $y'_x$\;
\caption{\tool}
\label{alg:algorithm}
\end{algorithm}

Algorithm~\ref{alg:algorithm} outlines the steps involved in the execution of {\tool}, our retrieval-augmented language model. The inputs to {\tool} include: an input context $x$, the original pre-trained model $\mathcal{M}$, the learning rate $\eta_{logits}$, the datastore $(K, V)$, the number of neighbors to retrieve $N$, and the number of iterative retrieval cycles $E$. 
The output generated by {\tool} is the updated prediction $y_x'$.

Initially, \tool computes the representation vector $r$ of the input $x$, using it to fetch the top-$N$ nearest neighbors from the datastore (lines~\ref{algo:getkey}–\ref{algo:getpair}). An iterative retrieval process then follows, which is a unique feature compared to existing methods. The iterative retrieval process is similar to the process of model fine-tuning conducted over a specified number of epochs, denoted as $E$ (lines~\ref{algo:startepoch}–\ref{algo:endepoch}). 
At each iteration, the original model's prediction (retrieved from $logits_x$ in line~\ref{algo:getlogits}) is adjusted based on the logits alteration computed from the retrieved neighbors (lines~\ref{alg:nblogitstart}–\ref{alg:nblogitend}).

It's worth noting that neighbors may vary in their relevance to the input context. Accordingly, we introduce weights $\lambda_i$ for each neighbor (line~\ref{algo:diffweights}). These weights are calculated based on the inverse of the distance between the neighbors and the input:
\begin{equation}
\label{eq:lambda}
\lambda_i = \frac{1/(d_i+1)}{\sum_{d\in D}(1/(d+1))}
\end{equation}

Intuitively, a smaller distance between a neighbor and the input results in a higher weight, meaning that closer neighbors contribute more substantially to the updated prediction.

Finally, \textit{\tool} updates the logits using the calculated change in logits, which has been interpolated from retrieved samples (line~\ref{algo:finallogit}). To facilitate further iterations of the retrieval process, the datastore is also updated (line~\ref{algo:updatestore}). While an ideal update method would involve recalculating the entire datastore using Equation~\ref{eq:finally}, we opt for a more computationally efficient strategy. Specifically, we maintain the same set of neighbors across all iterations and apply a constant $\Delta{logits}$ to the logits of these neighbors, balancing efficacy with computational efficiency.

\textbf{Discussions.} Differing from existing methods, {\tool} provides two main advantages. First, it employs detailed retrieval information, $\Delta {logits}$, for a more precise evaluation of each retrieved neighbor’s influence on the final prediction. Second, its iterative retrieval cycles can further improve performance. It's important to note that these multiple iterations are not actual fine-tuning, but rather a series of retrieval processes. These iterations are also optional and can be adjusted according to specific needs, like accuracy and inference efficiency. In our evaluation, we found that {\tool} outperformed baseline models even with just one iteration (the conventional setting). With multiple iterations, however, {\tool}'s performance can be further enhanced (see results in RQ4).

%% file: sections/setup.tex

The experimental design considers two completion scenarios: token-level and line-level completions, on models with or without fine-tuning. Specifically, we aim to answer the research questions:

\begin{itemize}[leftmargin=*]
\item \textbf{RQ1}: How effective is \tool in the two completion tasks?  

\item \textbf{RQ2}: To what extent can \tool approximate the effect of actual fine-tuning?


\item \textbf{RQ3}: How do different parameter settings, including the weighting strategies and the number of neighbours selected, affect {\tool}'s performance?

\item \textbf{RQ4}: How useful is the multi-round iteration strategy in {\tool}?
\end{itemize}

\subsection{Benchmarks}
\paragraph{Completion Scenario}
Based on the scale of completion, we consider two completion scenarios, i.e., token-level and line-level completions. For token-level completion, the model predicts the next token, based on the given (correct) context. The metric for evaluation in token-level completion is \textit{accuracy}, i.e., checking whether each completion is correct. 
\qi{For line-level completion, the model performs repeated execution of token-level completion until a line is completed, and retrieval occurs at every step of token prediction. }
Contrasting with token-level completions, predictions for each token depend on the prediction of the preceding token, which might be incorrect. In line with CodeXGLUE~\cite{lu2021codexglue}, the chosen evaluation metrics are \textit{exact match} (EM) and \textit{edit similarity} (ES).

\paragraph{Datasets}
We have chosen two widely used benchmarks for our study: the dataset from kNM-LM~\cite{tang2023domain} and the code completion benchmarks from CodeXGLUE~\cite{lu2021codexglue}.
Specifically, kNM-LM benchmark comprises 20 Java projects: 10 large-scale and 10 small-scale. In our experiments, we selected the ten larger projects.
CodeXGLUE benchmarks contain code samples written in both Java and Python programming languages, i.e., JavaCorpus~\cite{allamanis2013mining} and PY150~\cite{raychev2016probabilistic}.

We follow the settings in~\cite{tang2023domain, lu2021codexglue} for preparing and splitting the training and testing data. The training dataset can be used to fine-tune the pre-trained models. Note that the kNM-LM benchmarks do not provide a pre-defined test set tailored for line-level completion. To circumvent this, we follow the instructions in~\cite{allamanis2013mining}. Specifically, we randomly extract 300 lines of code from the test data of each project to serve as targets for model completion.
For evaluations regarding token-level predictions, we let the models predict each individual token in the test code samples.

\paragraph{Models}
Following the state-of-the-art work~\cite{tang2023domain}, we selected two widely used code pre-trained models in our experiments:  1) \textbf{CodeGPT}~\cite{lu2021codexglue}: It is a GPT-style code pre-trained model to support code completion. CodeGPT has the same model architecture and training objectives as GPT-2~\cite{radford2019language}, which consists of 12 layers of Transformer decoders. CodeGPT is pre-trained on Python and Java corpora from CodeSearchNet~\cite{husain2019codesearchnet}, which includes 1.1M Python code and 1.6M Java code. CodeGPT-adapted is pre-trained from GPT-2 and we use CodeGPT-small-java-adaptedGPT2 and CodeGPT-small-python-adaptedGPT2 to evaluate code completion in Java and Python datasets, respectively.
2) \textbf{UniXcoder}~\cite{guo2022unixcoder}: It is a cross-modal pre-trained model using mask attention matrices with prefix adapters to control the model behaviour. Furthermore, it leverages cross-modal contents such as AST and code comment to enhance the code representations. Specifically, it consists of 12 layers of Transformer with 768-dimensional hidden states. UniXcoder is pre-trained on the CodeSearchNet~\cite{husain2019codesearchnet} dataset for six programming languages including Java and Python. 

\begin{figure*}[!t]
\centering
\includegraphics[width=0.8\linewidth]{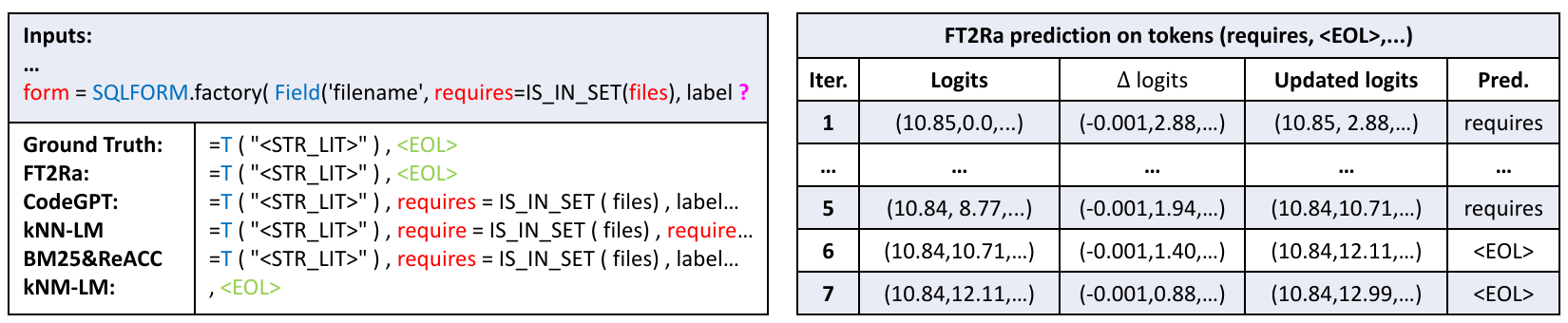}
\caption{Case study for CodeGPT line-level completion on PY150}
\label{fig:rq1-5}
\end{figure*}

Although our method is general, in this paper, we did not select very large models like CodeGen, InCoder, and CodeLLama, as fine-tuning them demands substantial computing resources. This requirement arises particularly because 1) the dataset includes unique symbols (e.g., $<STR\_LIT>$, $<NUM\_LIT>$, $<CHAR\_LIT>$) that necessitate specialized fine-tuning and 2) our experimental setting in RQ2 requires fine-tuning. Consequently, we chose two large CPMs, as suggested in the recent study~\cite{tang2023domain}.

\vspace{-2mm}

\subsection{Baselines}
Four state-of-the-art retrieval-based baselines are selected for comparisons, including kNN-LM, kNM-LM, BM25 and ReACC, where BM25 and ReACC are suitable to the line-level completions.
\begin{itemize}[leftmargin=*]
 \item \textbf{kNN-LM}~\cite{Khandelwal2020Generalization}: It augments the prediction of a pre-trained language model by linearly interpolating its next word distribution with a k-nearest neighbours model. The nearest neighbours are computed based on the distance in the vector space with a single forward pass of a pre-trained model over the retrieved dataset. The final distribution is the weighted sum of the original model distribution and the nearest neighbour distribution. 
    \item \textbf{kNM-LM}~\cite{tang2023domain}: It utilizes the in-domain code to construct the retrieved datastore decouple from LM and then combines with LM by Bayesian Inference for code completion. Compared with kNN-LM, it is able to calculate the posterior probability and utilize it to merge the distributions of nearest neighbours and the original model, which avoids manual weight tuning between the model distribution and neighbour distribution. 
    \item \textbf{BM25}~\cite{robertson2009probabilistic}: It is a term-based retrieval approach, which considers each code fragment as a code token sequence and employs bag-of-words representations to compute the matching score based on the lexical similarity between the query and document. Hence, it is more suitable for the line-level completion. As BM25 is based on the term frequency, it is one kind of sparse retriever.  
    \item \textbf{ReACC}~\cite{lu2022reacc}: It is a hybrid retriever framework by combining scores of sparse and dense retrievers. For sparse retrievers, it uses BM25~\cite{robertson2009probabilistic} for implementation. For dense retriever, it maps each code fragment to a dense vector based on the DPR model~\cite{karpukhin2020dense}, which consists of two bidirectional transformer encoders to encode the query code and the retrieved code for the retrieval.
   
\end{itemize}

\begin{table*}[!t]
\caption{Results of token-level completion on pre-trained models (\%).}
\footnotesize
\label{tb:rq1-1}
\resizebox{.87\textwidth}{!}{
\begin{tabular}{ccccccccccc}
\hline
\multirow{2}{*}{Type}                        & \multirow{2}{*}{Dataset} & \multicolumn{4}{c}{CodeGPT}                 &  & \multicolumn{4}{c}{UniXcoder}               \\ \cline{3-6} \cline{8-11} 
                                             &                          & Original & kNN-LM & kNM-LM & FT2Ra          &  & Original & kNN-LM & kNM-LM & FT2Ra          \\ \hline
\multicolumn{1}{c|}{\multirow{11}{*}{kNM-LM}} & Rest.                    & 46.99    & 54.99  & 70.71  & \textbf{77.68} &  & 42.59    & 50.86  & 71.26  & \textbf{77.58} \\
\multicolumn{1}{c|}{}                        & Amaze.                   & 55.22    & 58.33  & 66.34  & \textbf{71.00} &  & 54.65    & 56.73  & 68.14  & \textbf{71.79} \\
\multicolumn{1}{c|}{}                        & Dropwizard               & 50.11    & 55.00  & 65.50  & \textbf{71.14} &  & 47.15    & 51.30  & 66.56  & \textbf{70.12} \\
\multicolumn{1}{c|}{}                        & Eureka                   & 52.76    & 55.73  & 64.56  & \textbf{70.15} &  & 51.00    & 54.36  & 66.01  & \textbf{69.35} \\
\multicolumn{1}{c|}{}                        & Feign                    & 48.71    & 54.47  & 70.63  & \textbf{77.48} &  & 45.01    & 50.36  & 70.87  & \textbf{77.12} \\
\multicolumn{1}{c|}{}                        & Galaxy                   & 53.40    & 55.57  & 64.90  & \textbf{69.24} &  & 49.57    & 52.47  & 65.16  & \textbf{69.25} \\
\multicolumn{1}{c|}{}                        & Interview                & 64.70    & 66.46  & 69.29  & \textbf{73.14} &  & 62.91    & 64.80  & 71.54  & \textbf{75.27} \\
\multicolumn{1}{c|}{}                        & Logging.                 & 60.06    & 65.10  & 79.10  & \textbf{87.38} &  & 56.95    & 61.85  & 79.69  & \textbf{86.30} \\
\multicolumn{1}{c|}{}                        & Requery                  & 56.69    & 59.44  & 68.39  & \textbf{75.75} &  & 54.11    & 56.07  & 68.66  & \textbf{74.91} \\
\multicolumn{1}{c|}{}                        & Froyo.                   & 59.53    & 62.56  & 67.64  & \textbf{71.04} &  & 58.79    & 61.31  & 69.38  & \textbf{72.79} \\ \cline{2-11} 
\multicolumn{1}{c|}{}                        & Avg.                     & 54.82    & 58.77  & 68.71  & \textbf{74.40} &  & 52.27    & 56.01  & 69.73  & \textbf{74.45} \\ \hline
\multicolumn{1}{c|}{\multirow{2}{*}{CodeXGLUE}}  & JavaCorpus               & 64.95    & 67.83  & 70.74  & \textbf{72.07} &  & 65.61    & 67.92  & 72.13  & \textbf{74.73} \\
\multicolumn{1}{c|}{}                        & PY150                    & 52.41    & 55.35  & 60.94  & \textbf{62.19} &  & 60.44    & 64.62  & 69.81  & \textbf{71.43} \\ \hline
                                             & Total Avg.               & 55.46    & 59.24  & 68.23  & \textbf{73.19} &  & 54.07    & 57.72  & 69.93  & \textbf{74.22} \\ \hline
\end{tabular}
}
\end{table*}
\paragraph{Configurations}
Considering the hyper-parameters in Algorithm~\ref{alg:algorithm}, in our experiments, we configured the number of neighbors ($N$) and the number of iterations ($E$) to 20 and 7, respectively. Additionally, we set the learning rates ($\eta_{logits}$) to specific values: 3 for JavaCorpus, 5 for PY150, and 5 for the kNM-LM dataset. It's worth noting that we thoroughly evaluated and discussed various settings of these hyperparameters in RQ2 and RQ3.
For the other baseline methods, we selected the default configuration used in their papers.
Notably,  kNN-LM is not applied in code learning tasks, we followed the same configurations as described in~\cite{tang2023domain}.

%% file: sections/results.tex
\subsection{RQ1: Effectiveness on pre-trained models}



The main goal of the retrieval augmentation is to bolster the model's performance without the need for fine-tuning. Therefore, this experiment aims to evaluate the effectiveness of \tool on pre-trained models without fine-tuning.


\noindent \textbf{Token-level Completion.} The results for token-level completion are shown in Table~\ref{tb:rq1-1}, including the results on the ten Java projects from kNM-LM and the two CodeXGLUE benchmarks. The \textit{Original} column shows the results with the pre-trained models.

The overall results show that, compared with the original pre-trained models, all retrieval-augmented techniques have a higher accuracy, demonstrating the usefulness of retrieval-based augmentation. Furthermore, we can see that \tool significantly outperforms the baselines across all datasets and models. For example, while the average accuracies of pre-trained models on CodeGPT and UniXcoder stand at 55.46\% and 54.07\%, respectively, \tool increases the performance to 73.19\% and 74.22\%, outperforming all baseline models. Moreover, in comparison with the best baseline kNM-LM, \tool boasts an average increase of 4.96\% for CodeGPT and 4.29\%  for UniXcoder. The results demonstrate the effectiveness of \tool in token-level completion.

\begin{table*}[!t]
\caption{Results of line-level completion on pre-trained models (\%).}
\footnotesize
\label{tb:rq1-2}
\resizebox{.99\textwidth}{!}{
\begin{tabular}{ccccccccccccccccccc}
\hline
\multirow{3}{*}{Type}                           & \multirow{3}{*}{Dataset} & \multicolumn{17}{c}{CodeGPT}                                                                                                                                                                   \\ \cline{3-19} 
                                                &                          & \multicolumn{2}{c}{Original} &  & \multicolumn{2}{c}{kNN-LM} &  & \multicolumn{2}{c}{kNM-LM} &  & \multicolumn{2}{c}{BM25} &  & \multicolumn{2}{c}{ReACC} &  & \multicolumn{2}{c}{FT2Ra}       \\ \cline{3-4} \cline{6-7} \cline{9-10} \cline{12-13} \cline{15-16} \cline{18-19} 
                                                &                          & EM            & ES           &  & EM           & ES          &  & EM           & ES          &  & EM          & ES         &  & EM          & ES          &  & EM             & ES             \\ \hline
\multicolumn{1}{c|}{\multirow{11}{*}{kNM-LM}}   & Rest.                    & 1.00          & 49.36        &  & 1.00         & 53.68       &  & 9.63         & 47.05       &  & 3.99        & 53.48      &  & 3.99        & 53.05       &  & \textbf{17.94} & \textbf{70.62} \\
\multicolumn{1}{c|}{}                           & Amaze.                   & 1.99          & 56.86        &  & 3.99         & 57.94       &  & 9.30         & 47.75       &  & 2.99        & 59.25      &  & 2.99        & 58.89       &  & \textbf{22.92} & \textbf{66.54} \\
\multicolumn{1}{c|}{}                           & Dropwizard               & 1.00          & 52.13        &  & 2.33         & 57.83       &  & 4.65         & 49.56       &  & 1.33        & 52.52      &  & 1.33        & 51.84       &  & \textbf{22.59} & \textbf{69.34} \\
\multicolumn{1}{c|}{}                           & Eureka                   & 3.99          & 55.76        &  & 5.32         & 58.20       &  & 8.31         & 50.41       &  & 3.99        & 57.59      &  & 3.99        & 57.78       &  & \textbf{20.93} & \textbf{68.00} \\
\multicolumn{1}{c|}{}                           & Feign                    & 1.33          & 47.72        &  & 3.32         & 52.77       &  & 10.63        & 47.19       &  & 3.99        & 53.50      &  & 3.99        & 53.21       &  & \textbf{25.91} & \textbf{72.34} \\
\multicolumn{1}{c|}{}                           & Galaxy                   & 1.33          & 50.97        &  & 2.66         & 54.61       &  & 11.96        & 48.53       &  & 2.66        & 51.72      &  & 2.33        & 52.07       &  & \textbf{22.26} & \textbf{64.14} \\
\multicolumn{1}{c|}{}                           & Interview                & 8.97          & 61.63        &  & 13.95        & 62.80       &  & 19.60        & 57.28       &  & 9.30        & 61.41      &  & 9.97        & 62.91       &  & \textbf{27.91} & \textbf{71.08} \\
\multicolumn{1}{c|}{}                           & Logging.                 & 2.66          & 59.04        &  & 5.98         & 63.13       &  & 15.28        & 58.59       &  & 6.31        & 63.42      &  & 6.64        & 63.64       &  & \textbf{33.89} & \textbf{80.63} \\
\multicolumn{1}{c|}{}                           & Requery                  & 5.98          & 61.76        &  & 8.97         & 62.81       &  & 9.63         & 46.51       &  & 6.64        & 63.21      &  & 7.31        & 63.12       &  & \textbf{28.24} & \textbf{73.11} \\
\multicolumn{1}{c|}{}                           & Froyo.                   & 8.31          & 63.96        &  & 11.30        & 65.85       &  & 16.28        & 55.58       &  & 7.97        & 63.90      &  & 8.31        & 63.93       &  & \textbf{28.24} & \textbf{73.91} \\ \cline{2-19} 
\multicolumn{1}{c|}{}                           & Avg.                     & 3.65          & 55.92        &  & 5.88         & 58.96       &  & 11.53        & 50.85       &  & 4.92        & 58.00      &  & 5.08        & 58.04       &  & \textbf{25.08} & \textbf{70.97} \\ \hline
\multicolumn{1}{c|}{\multirow{2}{*}{CodeXGLUE}} & JavaCorpus               & 13.69         & 48.74        &  & 16.28        & 49.90       &  & 16.18        & 43.89       &  & 14.19       & 49.73      &  & 14.19       & 49.85       &  & \textbf{22.88} & \textbf{54.54} \\
\multicolumn{1}{c|}{}                           & PY150                    & 0.00          & 12.54        &  & 10.39        & 41.40       &  & 8.69         & 11.56       &  & 0.20        & 13.35      &  & 0.20        & 13.36       &  & \textbf{18.48} & \textbf{50.52} \\ \hline
                                                & Total Avg.               & 4.19          & 51.71        &  & 7.12         & 56.74       &  & 11.68        & 46.99       &  & 5.30        & 53.59      &  & 5.43        & 53.64       &  & \textbf{24.35} & \textbf{67.90} \\ \hline
\multirow{3}{*}{Type}                           & \multirow{3}{*}{Dataset} & \multicolumn{17}{c}{UniXcoder}                                                                                                                                                                 \\ \cline{3-19} 
                                                &                          & \multicolumn{2}{c}{Original} &  & \multicolumn{2}{c}{kNN-LM} &  & \multicolumn{2}{c}{kNM-LM} &  & \multicolumn{2}{c}{BM25} &  & \multicolumn{2}{c}{ReACC} &  & \multicolumn{2}{c}{FT2Ra}       \\ \cline{3-4} \cline{6-7} \cline{9-10} \cline{12-13} \cline{15-16} \cline{18-19} 
                                                &                          & EM            & ES           &  & EM           & ES          &  & EM           & ES          &  & EM          & ES         &  & EM          & ES          &  & EM             & ES             \\ \hline
\multicolumn{1}{c|}{\multirow{11}{*}{kNM-LM}}   & Rest.                    & 0.66          & 50.12        &  & 1.66         & 52.07       &  & 11.30        & 54.05       &  & 1.99        & 63.46      &  & 1.99        & 64.66       &  & \textbf{18.94} & \textbf{72.17} \\
\multicolumn{1}{c|}{}                           & Amaze.                   & 1.33          & 56.00        &  & 1.66         & 58.40       &  & 13.29        & 55.28       &  & 1.00        & 58.87      &  & 1.00        & 59.04       &  & \textbf{23.26} & \textbf{66.92} \\
\multicolumn{1}{c|}{}                           & Dropwizard               & 0.00          & 49.44        &  & 1.33         & 54.89       &  & 16.61        & 56.41       &  & 0.66        & 59.82      &  & 0.66        & 58.33       &  & \textbf{20.27} & \textbf{69.94} \\
\multicolumn{1}{c|}{}                           & Eureka                   & 0.33          & 55.44        &  & 1.66         & 59.39       &  & 15.61        & 58.91       &  & 0.00        & 63.56      &  & 0.00        & 62.23       &  & \textbf{22.26} & \textbf{69.98} \\
\multicolumn{1}{c|}{}                           & Feign                    & 0.00          & 50.06        &  & 1.00         & 52.46       &  & 8.97         & 52.99       &  & 4.65        & 67.68      &  & 4.65        & 67.39       &  & \textbf{27.24} & \textbf{75.24} \\
\multicolumn{1}{c|}{}                           & Galaxy                   & 0.33          & 50.06        &  & 0.33         & 53.06       &  & 13.29        & 47.61       &  & 1.00        & 54.89      &  & 1.00        & 54.86       &  & \textbf{21.26} & \textbf{64.56} \\
\multicolumn{1}{c|}{}                           & Interview                & 4.65          & 59.89        &  & 7.31         & 61.88       &  & 20.93        & 62.78       &  & 2.66        & 58.28      &  & 3.32        & 57.65       &  & \textbf{30.56} & \textbf{73.34} \\
\multicolumn{1}{c|}{}                           & Logging.                 & 0.00          & 55.11        &  & 4.98         & 60.78       &  & 17.28        & 59.45       &  & 3.32        & 72.28      &  & 3.99        & 72.45       &  & \textbf{35.22} & \textbf{81.23} \\
\multicolumn{1}{c|}{}                           & Requery                  & 2.33          & 60.56        &  & 3.99         & 61.92       &  & 14.95        & 56.73       &  & 2.33        & 65.34      &  & 2.33        & 65.14       &  & \textbf{30.56} & \textbf{73.79} \\
\multicolumn{1}{c|}{}                           & Froyo.                   & 1.33          & 63.60        &  & 1.99         & 65.14       &  & 21.93        & 62.75       &  & 2.33        & 63.94      &  & 2.66        & 64.09       &  & \textbf{32.56} & \textbf{76.46} \\ \cline{2-19} 
\multicolumn{1}{c|}{}                           & Avg.                     & 1.10          & 55.03        &  & 2.59         & 58.00       &  & 15.42        & 56.70       &  & 1.99        & 62.81      &  & 2.16        & 62.58       &  & \textbf{26.21} & \textbf{72.36} \\ \hline
\multicolumn{1}{c|}{\multirow{2}{*}{CodeXGLUE}} & JavaCorpus               & 8.49          & 47.72        &  & 9.99         & 49.69       &  & 12.89        & 48.06       &  & 7.79        & 46.97      &  & 7.79        & 47.09       &  & \textbf{24.58} & \textbf{58.67} \\
\multicolumn{1}{c|}{}                           & PY150                    & 0.10          & 8.43         &  & 0.00         & 8.60        &  & 0.10         & 12.74       &  & 0.00        & 7.46       &  & 0.00        & 7.43        &  & \textbf{29.17} & \textbf{59.00} \\ \hline
                                                & Total Avg.               & 1.63          & 50.54        &  & 2.99         & 53.19       &  & 13.93        & 52.31       &  & 2.31        & 56.88      &  & 2.45        & 56.70       &  & \textbf{26.32} & \textbf{70.11} \\ \hline
\end{tabular}
}
\end{table*}

\noindent \textbf{Line-level Completion.}
The results for line-level completion, evaluated on CodeGPT and UniXcoder, are presented in Table~\ref{tb:rq1-2}. The metrics of exact match and edit similarity are represented by the columns \textit{EM} and \textit{ES}, respectively. 
Similarly, we can find that all of the retrieval-based methods could still enhance the performance, but the improvement degree of baselines is generally limited. For example, on average, kNN-LM, kNM-KM, BM25 and ReACC achieve scores (7.12, 56.74), (11.68, 46.99), (5.30, 53.59) and (5.43, 53.64) on CodeGPT, respectively, while the pre-trained model achieves (4.19, 51.71). 
While the recent state-of-the-art kNM-LM can achieve higher EM scores than other baselines, its ES scores are lower.
The low performance of baselines could be attributed to the difficulty of the line-level completion. Any incorrect token prediction (inaccurate context) could affect the prediction of the following tokens. 
It is obvious that \tool significantly outperforms the baselines, manifesting its superiority in both the EM and ES metrics across all datasets and models. Considering the results on CodeGPT, \tool increases the scores to (24.35, 67.90). While on UniXcoder, \tool achieves higher improvement (26.32, 70.11) than the pre-trained model (1.63, 50.54) and the baselines. The results demonstrate the effectiveness of our proposed retrieval mechanism on line-level completion.

We have observed that the performance of various methods varies across different datasets and models. Considering the results on the dataset {Froyo.}, we find that all methods consistently achieve higher EM scores on CodeGPT compared to UniXcoder. Interestingly, all baseline models, including pre-trained ones, exhibit poor performance on the PY150 dataset but demonstrate better results on the {JavaCorpus} dataset. 
Upon our in-depth analysis of CodeGPT, we discovered that the pre-trained model CodeGPT-small-py-adaptedGPT2 tends to underestimate the probability of end-of-line tokens (<EOL>). We randomly selected 30 test data instances from PY150, specifically targeting cases where \tool succeeded while the original model failed. We discovered that 9 of these instances reached the maximum token prediction count (set at 100) when predicted by CodeGPT.
In contrast, we randomly checked 100 instances in JavaCorpus predicted by CodeGPT-small-java-adaptedGPT2, and none of the predictions reached the token count limit. This discrepancy could be attributed to the natural line termination indicators present in Java code such as semicolons and braces, which allow the model to easily discern when to stop the prediction. However, in Python, the model must accurately predict the <EOL> symbol to recognize the end of a statement. Compared with others, \tool exhibits significant enhancements on the {PY150} dataset, with improvements of (18.48, 50.52) and (29.17, 59.00) when evaluated on CodeGPT and UniXcoder, respectively.


\begin{figure}[!t]
\centering
\includegraphics[width=1\linewidth]{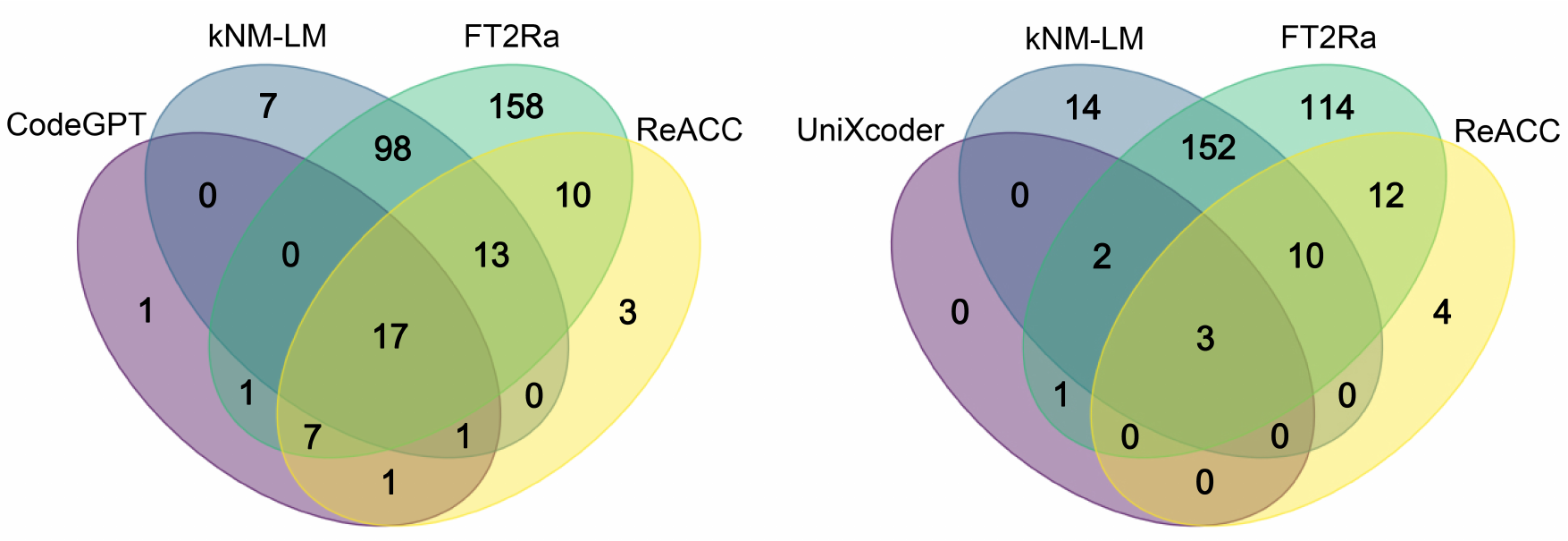}
\caption{Venn diagram of the EM results on CodeGPT (left) and UniXcoder (right).}
\label{fig:rq1-4}
\vspace{-4mm}
\end{figure}

In Figure~\ref{fig:rq1-5}, we present an illustrative example that shows the advantage of \tool when applied to PY150. Upon examining the results obtained by different methods, we observe that, except for kNM-LM, all models accurately predict the initial seven tokens. However, when reaching the eighth token, the baselines, including the original model, cannot predict the correct termination token <EOL>. Instead, they predict the token \textit{requires}, which leads to an uninterrupted sequence of predictions until reaching the maximum token count. By checking CodeGPT's prediction on the eighth token, we discover that the token \textit{requires} has the highest prediction probability (0.13), while the token <EOL> receives a prediction probability of 0, making it challenging for the baselines to correct the prediction. kNM-LM exhibits too much augmentation, resulting in incorrect predictions for even the first token.
The table on the right provides detailed insights into how \tool corrects the prediction. Despite the stubborn prediction of \textit{requires}, \tool leverages the calculation of $\Delta_{logits}$ across multiple iterations to steadily increase the logits of the token <EOL> while decreasing the logits of \textit{requires}. Ultimately, at the sixth iteration, \tool successfully fixes the prediction.

In Figure~\ref{fig:rq1-4}, we present a Venn diagram depicting the completion lines that achieve an exact match with the ground truth. For the sake of clarity, we have excluded the results of BM25 and kNN-LM from the diagram since their outcomes closely resemble those of ReACC and kNM-LM. The findings clearly illustrate that \tool outperforms other methods by generating a significantly larger number of unique code lines.

\begin{table}[!t]
\centering
\caption{Results of average generation time per token (s).}
\footnotesize
\label{tb:rq1-3}
\begin{tabular}{cccccccc}
\hline
          &          & \multicolumn{2}{c}{Input Retrieval} &  & \multicolumn{3}{c}{Output Retrieval} \\ \cline{3-4} \cline{6-8} 
          & Original & BM25             & ReACC            &  & kNN-LM     & kNM-LM     & FT2Ra      \\ \hline
CodeGPT   & 0.0163   & 0.0161           & 0.0164           &  & 0.0208     & 0.0193     & 0.0271     \\
UniXcoder & 0.0134   & 0.0143           & 0.0135           &  & 0.0163     & 0.0155     & 0.0214     \\ \hline
\end{tabular}
\vspace{-4mm}
\end{table}

\noindent \textbf{Performance.} 
To evaluate \tool's performance, we measured the average time required to predict a token. We did not compare the line prediction time
since the predictions of different methods can have different lengths. Specifically, we selected 1,000 line-completion tasks at random, using different methods to predict the line with a set length of 100 tokens. We then recorded the average token prediction time for comparison. All experiments were conducted on a single A5000 GPU card for consistency.

Table~\ref{tb:rq1-3} presents the results. Note that the time used by \tool is from its seven retrieval iterations. On the CodeGPT model, the average prediction times for the original model, BM25, ReACC, kNN-LM, kNM-LM, and \tool are 0.0163s, 0.0161s, 0.0164s, 0.0208s, 0.0193s, and 0.0271s, respectively, and a similar trend is observed with UniXcoder. The results indicate that while input retrieval methods slightly impact prediction speed, output retrieval methods, which require more computation, tend to slow it down more noticeably. Compared to other output retrieval baselines, \tool, which retrieves more detailed information and allows for multiple retrieval rounds, takes slightly longer. This represents a trade-off between effectiveness and efficiency, with \tool sacrificing some speed for significant improvements in effectiveness.

\vspace{5pt}\noindent \fbox{
\parbox{0.95\linewidth}{\textbf{Answers to RQ1}: The results reveal \tool's dominant performance on pre-trained models over existing baselines in both token-level and line-level completions.}
}

\subsection{RQ2: Comparison with Fine-tuning}
The key insight of \tool lies an innovative approach that emulates the fine-tuning process with certain approximations (refer to Section~\ref{sec:inspiration}). Hence, we compared the results of \tool with genuine fine-tuning results on the kNM-LM dataset. 
We utilized the training data from all ten projects to fine-tune the pre-trained models and subsequently evaluated the methodologies on all test data of these projects. The pre-trained models were fine-tuned over a range of epochs. We compared the performance of various methods for both line-level and token-level completion, with the models fine-tuned across these different epochs.

\begin{figure}[!t]
\centering
\includegraphics[width=1\linewidth]{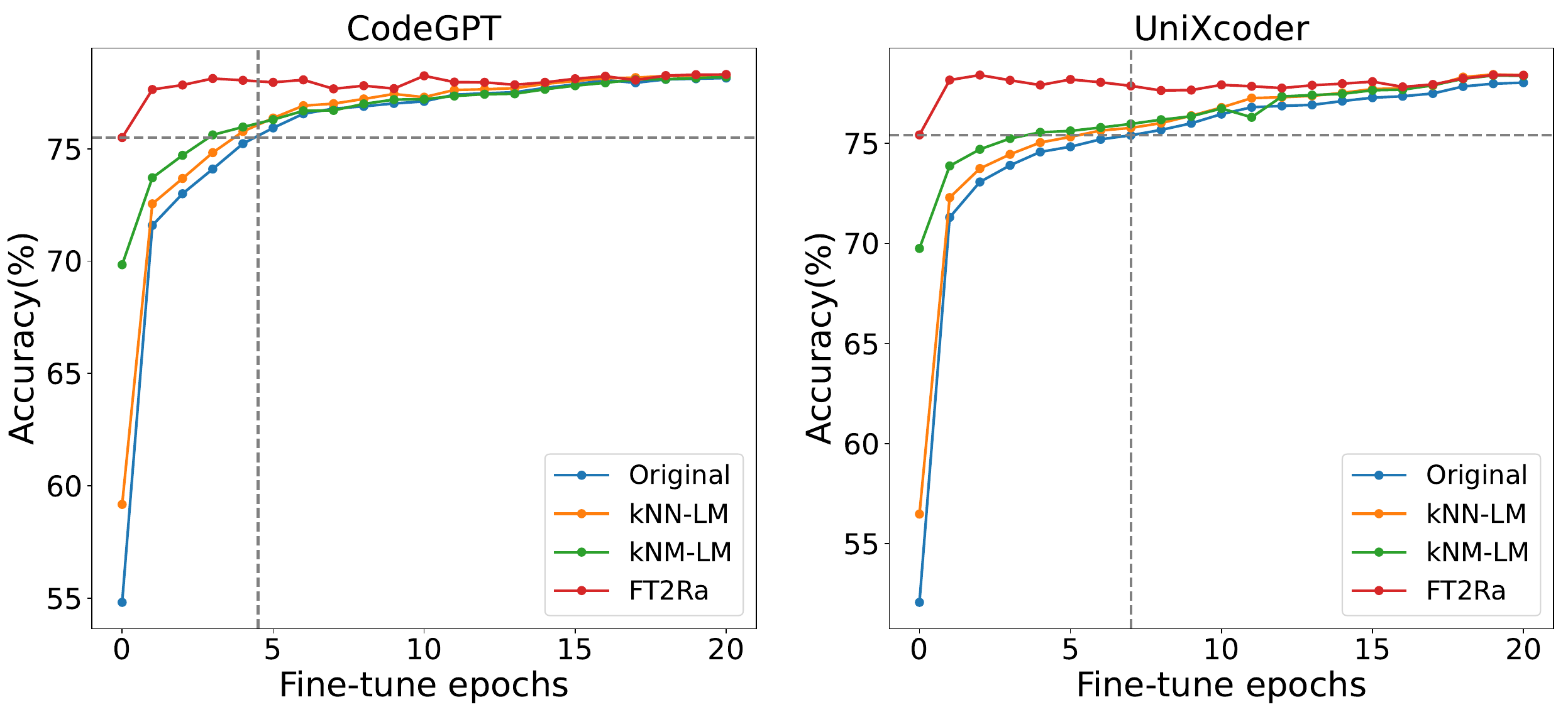}
\caption{Results of token-level completion on fine-tuned models with different epochs.}
\label{fig:rq1-tokenfine}
\vspace{-4mm}
\end{figure}
\noindent\textbf{Token-level Completion.} 
For the token-level completion, we capped the maximum number of epochs at 20. This upper limit was chosen because it was observed that most methods tend to reach convergence within this period.
Figure~\ref{fig:rq1-tokenfine} presents the token-level completion performance of different methods on fine-tuned models over various epochs. Notably, the comparative results at each point are derived by the retrieval from the respective fine-tuned models at specific epochs (i.e., epochs 1, 2, \ldots, 20).
The data stores are also updated under different fine-tuned models. The results for epoch 0 are derived from the pre-trained model without fine-tuning.

Overall, we observe a progressive improvement in the performance of the original model with an increasing number of fine-tuning epochs (see blue lines). The results of the retrieval-based methods also exhibit an upward trend, showing the generalization capability across different fine-tuned models. However, as the model goes through multiple fine-tuning epochs, the improvements are diminishing as the model nears its best performance after sufficient tuning.
Comparing \tool to the baselines, it is clear that \tool consistently outperforms the baselines on fine-tuned models.

To assess how closely \tool's effect (simulating fine-tuning) aligns with real fine-tuning, we compare \tool's performance on the pre-trained model without any fine-tuning to that of the fine-tuned models. As indicated by the dotted line, \tool, without fine-tuning the model, achieves similar performance to fine-tuned CodeGPT and UniXcoder models after approximately 4 and 7 epochs, respectively. In contrast, the best baseline, kNM-LM, only reaches a similar performance level with a model fine-tuned for about one epoch. These results underscore the value of our theoretical analysis from the fine-tuning process.

\begin{figure}[!t]
\centering
\includegraphics[width=1\linewidth]{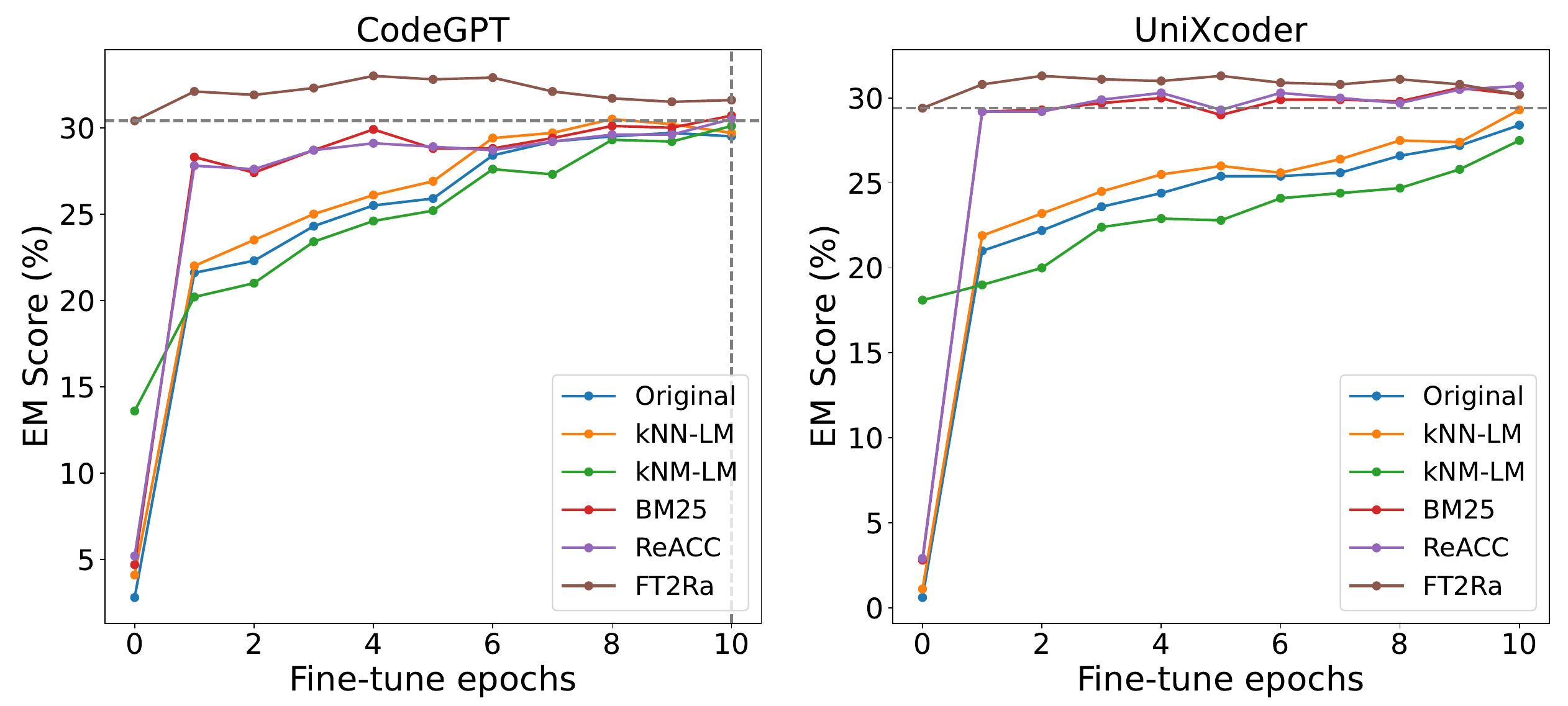}
\caption{Results of line-level completion.}
\label{fig:rq1-linefine}
\vspace{-2mm}
\end{figure}

\noindent\textbf{Line-level Completion.} 
Figure~\ref{fig:rq1-linefine} illustrates the results in terms of EM for line-level completion. Due to space constraints, the results for Edit Similarity can be accessed on our website~\cite{ft2rawebsite}. We capped the maximum number of epochs at 10 due to the large computational overhead of line-based completion.
When compared to the token-level completion results in Figure~\ref{fig:rq1-tokenfine}, it becomes evident that the impact of other baseline methods is notably diminished in line-level completion, primarily because this task is more difficult.
We observe that BM25 and ReACC yield similar results, likely due to their adoption of similar methods. On the other hand, the performance of kNN-LM and kNM-LM is very close to that of the fine-tuned models, which indicates that they have limited improvement.

Conversely, \tool continues to demonstrate clear advantages over other methods, due to its precise token prediction. Notably, when comparing the performance of \tool at epoch 0 with that of other fine-tuned models, it becomes apparent that even without fine-tuning, FT2Ra can
outperform the performance of fine-tuned models at 10 epochs.


\vspace{5pt}\noindent \fbox{
\parbox{0.95\linewidth}{\textbf{Answers to RQ2}: 
\tool remains highly effective when applied to fine-tuned models. Furthermore, our results indicate that \tool yields promising outcomes even without fine-tuning, achieving competitive or superior performance compared to fine-tuned models with multiple epochs.}
}

\begin{table}[!t]
\centering
\caption{Results with different weighting strategies and different numbers of neighbors (\%).}
\label{tb:rq3-1}
\resizebox{.45\textwidth}{!}{
\begin{tabular}{cccccccccc}
\hline
\multirow{2}{*}{Dataset} & \multicolumn{4}{c}{Weighting Strategy}          &  & \multicolumn{4}{c}{\#Neighbors}                          \\ \cline{2-5} \cline{7-10} 
                         & Rec.           & Uni.  & Smax  & Smax-T         &  & 5              & 10    & 20             & 50             \\ \hline
Rest.                    & 78.15          & 74.62 & 78.43 & \textbf{78.46} &  & \textbf{78.69} & 78.65 & 78.15          & 76.79          \\
Eureka                   & \textbf{70.65} & 68.80 & 67.12 & 66.84          &  & 69.38          & 70.14 & \textbf{70.65} & 69.79          \\
JavaCorpus               & \textbf{72.07} & 71.97 & 67.98 & 68.15          &  & 70.58          & 71.40 & 72.07          & \textbf{72.46} \\
PY150                    & \textbf{62.19} & 61.17 & 56.85 & 56.94          &  & 60.93          & 61.69 & \textbf{62.19} & 61.93          \\ \hline
\end{tabular}
}
\vspace{-4mm}
\end{table}


\subsection{RQ3: Impact of Weighting Strategy and the Number of Neighbors}

To evaluate the effectiveness of our weighting strategy and to understand the impact of the number of neighbours, we collected four datasets, including the two Java projects that were randomly chosen from the kNM-LM dataset, and the two datasets from CodeXGLUE. The evaluation is performed on the token-level completion task, which serves as the foundation for line-level completion.

\subsubsection{Effectiveness of the weighting strategy}

Since various baseline methods employ different weighting strategies to determine the significance of the retrieved samples. For instance, kNN-LM utilizes the softmax (referred to as \textit{Smax}), whereas kNM-LM employs softmax with temperate (denoted as \textit{Smax-T}). Our method calculates weights based on distance, referred to as \textit{Rec.} (see Equation~\ref{eq:lambda}). To provide a comparative evaluation, we incorporated these strategies into \tool for the comparisons. Additionally, we introduced a baseline, i.e., uniform strategy (\textit{Uni.}), which allocates equal weights to all samples.
Detailed results can be found on the left of Table~\ref{tb:rq3-1}. Obviously, the weighting strategy \textit{Rec.} outperforms other strategies when they are adopted to \tool. An exception is the results on Rest., where \textit{Smax} and \textit{Smax-T} marginally exceed the performance of \tool. Interestingly, the uniform strategy \textit{Uni.} excels over the other two methods for the benchmark JavaCorpus and PY150, emphasizing the importance of designing a suitable weighting strategy.



\subsubsection{Impact of the number of neighbours}
We evaluated the performance of \tool by setting the number of neighbors to 5, 10, 20, and 50. The findings, as presented in the right part of Table~\ref{tb:rq3-1}, suggest that \tool exhibits relatively limited sensitivity to the number of neighbours chosen. There is not a single optimal parameter that is universally effective across all datasets. In general, selecting too few neighbours may not provide enough information to augment predictions. Conversely, selecting an excessive number might introduce negative effects, such as the interference of irrelevant neighbours.

\vspace{5pt}\noindent \fbox{
\parbox{0.95\linewidth}{\textbf{Answers to RQ3}: Our weighting strategy is useful in enhancing the performance of \tool. Moreover, \tool generally exhibits limited sensitivity to changes in the number of neighbours.}
}

\begin{figure}[!t]
\centering
\includegraphics[width=0.95\linewidth]{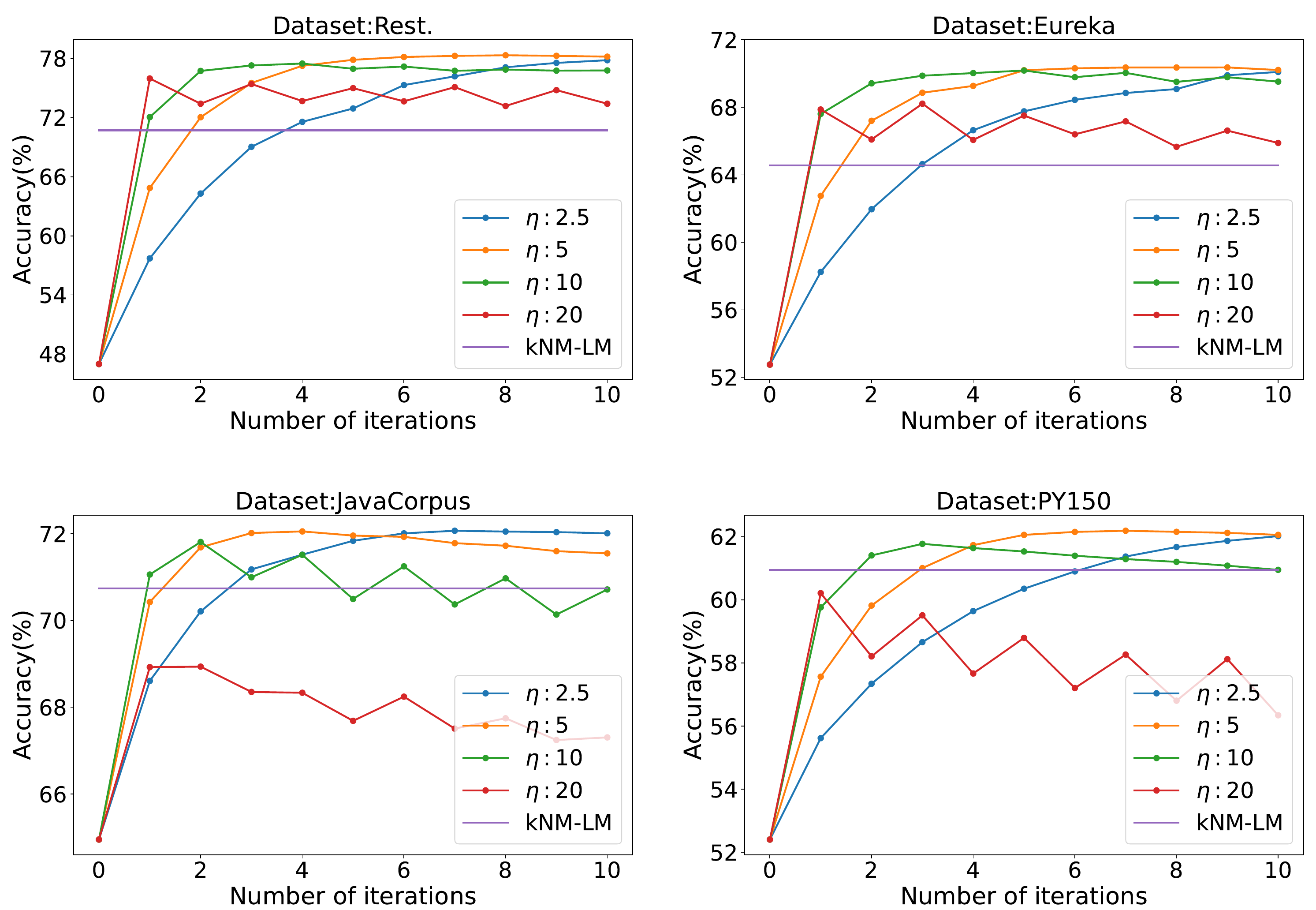}
\caption{Results with different numbers of iterations.}
\label{fig:rq4-1}
\vspace{-4mm}
\end{figure}

\subsection{RQ4: Usefulness of Multiple Iterations}
To evaluate the effect of the multiple iteration strategy incorporated into \tool, we configured \tool with varying retrieval iterations (i.e., $E$ in Algo.~\ref{alg:algorithm}) ranging from 1 to 10. We also consider the impact of the parameter $\eta_{logits}$, which are configured with 4 values: 2.5, 5, 10, and 20.
Evaluations were carried out using similar configurations as in RQ3, i.e., token-level completion on pre-trained models.

The results are presented in Figure~\ref{fig:rq4-1}. It is obvious that \tool's performance benefits from multiple retrieval rounds, which is a unique feature compared to existing retrieval-based baselines. By increasing the number of retrieval rounds, the performance of \tool gradually gets better. From the results, we found that the performance of \tool tends to stabilize after approximately 4 retrieval iteration cycles.

With respect to different \textit{learning rates} (i.e., $\eta_{logits}$), \tool's performance shows high sensitivity to this parameter. In general, larger values of $\eta_{logits}$ enable \tool to converge faster, whereas smaller ones necessitate multiple iterations. For instance, with $\eta_{logits}$ set to 0.5, convergence tends to be achieved after 10 iterations. In contrast, a setting of 4 for $\eta_{logits}$ reaches optimal performance after just one iteration. Yet, we also observed that excessively high learning rates could hamper \tool's performance. For instance, settings of $\eta_{logits}$ at 2 and 4 typically yield results inferior to those achieved with 0.5 and 1. The configuration using a value of 4 for $\eta_{logits}$ achieves the poorest performance. 
The \textit{learning rate} in \tool shows a similar effect to the learning rate of real training.

Even with just one iteration, \tool surpasses the best-performing baseline, kNM-LM, as shown by the straight line in Figure~\ref{fig:rq4-1}. A higher learning rate, (e.g., $\eta_{logits}=10$), is typically needed for faster convergence. Under this setting, \tool outperforms kNM-LM in Rest., EureKa, and JavaCorpus. In PY 150, \tool exceeds kNM-LM's performance after only 2 epochs. These results further highlight \tool's effectiveness, even with no or a few iterations.

While automatically selecting optimal parameters for learning rate and number of iterations can be challenging, there are some general guidelines that can aid in this process. Conducting initial trials on a small test set allows for the assessment of the model's performance. If the model exhibits rapid oscillation and a decline in performance, it suggests that the learning rate is too high. Conversely, if the model fails to converge after many iterations, it implies that the learning rate is too low. To adjust the number of iterations, early stopping techniques can be employed, ensuring that the tuning process is both computationally efficient and completed within a reasonable timeframe.


\vspace{5pt}\noindent \fbox{
\parbox{0.95\linewidth}{\textbf{Answers to RQ4}: 
 The multiple iteration strategy is useful in improving \tool's performance. In general, the more rounds, the better the results. Moreover, \tool shows sensitivity to the learning rate parameter, $\eta_{logits}$. Smaller values tend to yield superior results, but they require a greater number of iterations for convergence. }
}

%% file: sections/discussion.tex
Potential biases from our choices of models and datasets represent a possible threat to our study. To mitigate it, we have followed the recent works~\cite{tang2023domain} for guidance and selected two prominent datasets, i.e., the kNM-LM datasets and the CodeXGLUE benchmarks, and two widely-used pre-tained models, specifically CodeGPT and UniXcoder. We also plan to evaluate \tool on the large models such as CodeLLama and CodeGen in future work.
Furthermore, we were unable to establish a concrete theoretical framework to determine the weighting strategy. Instead, we empirically evaluated four common strategies in RQ3 and selected the most effective one. We acknowledge the significance of the weighting strategy and intend to investigate it further in future research. Another potential threat to our study is that the approximations inherent in retrieval-augmented methods may affect the precision of the results. This is particularly relevant when applying these methods to new models or datasets, where the impact of approximations might be more pronounced.
In line with~\cite{tang2023domain}, there is a threat to the use of ReACC on Java programs. The original authors only made their retrieval models for Python available, leaving the Java version undisclosed. To circumnavigate this obstacle during our Java experiments, we utilized their Python version. In parallel, we incorporated the BM25 model, which has a similar performance to ReACC.
For transparency, we have made our entire codebase, datasets, and experimental results public, thereby enabling independent verification.

%% file: sections/related.tex
\textbf{Code Completion.}
Code completion is regarded as a vital aspect of enhancing software development efficiency in contemporary Integrated Development Environments (IDEs). Hindle~\cite{hindle2016naturalness} were pioneers in employing N-gram techniques to implement code completion using language models. Subsequently, deep neural networks~\cite{liu2016neural} and pre-training techniques~\cite{feng2020codebert,wang2021codet5,liu2023contrabert, liu2022commitbart} have been made great progress.
While some of these efforts involve encoding code-specific structured information like Abstract Syntax Tree (AST) into inputs~\cite{li2017code,kim2021code}, the prevailing trend in current research treats source code as sequences of code tokens, as exemplified by models like CodeGPT~\cite{lu2021codexglue}, and UniXcoder~\cite{guo2022unixcoder}. The advent of large language models like ChatGPT~\cite{ChatGPTblog}, CodeGen~\cite{nijkamp2022codegen}, StarCoder~\cite{li2023starcoder} has introduced new opportunities and challenges to code completion. 
large models entail a vast number of parameters, which significantly elevates the cost of fine-tuning. Therefore, research on retrieval-based enhancement is essential in this context.

\noindent \textbf{Retrieval-augment Language Model.}
Retrieval-augmented techniques~\cite{shi2023replug,jiang2023active,he2021efficient, liu2020retrieval, liu2020atom} are primarily categorized into two types: one being retrieval enhancement applied to inputs, also referred to as pre-task retrieval. This category encompasses techniques such as REACC~\cite{lu2022reacc}, REDCODER~\cite{parvez2021retrieval} and DPR~\cite{karpukhin2020dense} as discussed in previous works. These retrieval techniques necessitate the preliminary segmentation of the data to be retrieved into fixed-length chunks, with each chunk typically containing several hundred tokens. They concatenated the most relevant information to the inputs for the enhancement. 
\qi{Some works go beyond simply retrieving from the original training set, they refine new information from the original training dataset. ASAP~\cite{ahmed2024automatic}, in the task of code summarization, uses not just the conventional source code and summary as input but also incorporates static analysis products such as the repository name, the fully qualified name of the target function, its signature and its data flow graph. RLPG~\cite{shrivastava2023repository} is proposed for single-line code auto-completion in an IDE. RLPG not only retrieves similar content as supplemental input but also utilizes repository-level code context such as Post Lines, Identifiers, Type Identifiers as additional input prompts. Joshi et al.~\cite{joshi2023repair} proposes a multi-lingual program repair method named RING. It retrieves relevant buggy-fix examples from an example bank, using the completed bug repair and repair methods as supplemental input prompts.}
On the other hand, some works also try to focus on retrieval enhancement for outputs such as kNN-LM~\cite{Khandelwal2020Generalization,xu2023nearest}, kNM-LM~\cite{tang2023domain}, and RETRO~\cite{borgeaud2022improving}. This kind of retrieval paradigm involves the preliminary creation of a retrieval database, where information from this database is utilized to modify the output generated. For example, RETRO~\cite{borgeaud2022improving} integrates it within the Transformer, while kNN-LM~\cite{Khandelwal2020Generalization,xu2023nearest} employ probability interpolation at the final probability layer. Compared with these works, we develop a novel retrieval-based approach from theoretical analysis to mimic genuine fine-tuning for code completion.

%% file: sections/conclusion.tex
In this paper, we introduce a novel retrieval augmentation method for code completion tasks. 
Guided by a theoretical analysis, we discerned the value of $\Delta logits$ as a pivotal retrieval metric. Building on this revelation, we designed \tool, a method that is to simulate the fine-tuning process closely. Similarly, \tool incorporates a learning rate and a multi-round iteration strategy, aiming to mirror the results of genuine fine-tuning.
The experimental results demonstrated \tool's superiority against state-of-the-art methods and its competitive results with regards to fine-tuning.

%% file: sample-acmsmall-conf.bbl

\begin{thebibliography}{54}


\ifx \showCODEN    \undefined \def \showCODEN     #1{\unskip}     \fi
\ifx \showDOI      \undefined \def \showDOI       #1{#1}\fi
\ifx \showISBNx    \undefined \def \showISBNx     #1{\unskip}     \fi
\ifx \showISBNxiii \undefined \def \showISBNxiii  #1{\unskip}     \fi
\ifx \showISSN     \undefined \def \showISSN      #1{\unskip}     \fi
\ifx \showLCCN     \undefined \def \showLCCN      #1{\unskip}     \fi
\ifx \shownote     \undefined \def \shownote      #1{#1}          \fi
\ifx \showarticletitle \undefined \def \showarticletitle #1{#1}   \fi
\ifx \showURL      \undefined \def \showURL       {\relax}        \fi
\providecommand\bibfield[2]{#2}
\providecommand\bibinfo[2]{#2}
\providecommand\natexlab[1]{#1}
\providecommand\showeprint[2][]{arXiv:#2}

\bibitem[git(2022)]%
        {githubcopilot}
 \bibinfo{year}{2022}\natexlab{}.
\newblock \bibinfo{title}{{GitHub Copilot}}.
\newblock
\newblock
\newblock
\shownote{\url{https://github.com/features/copilot}}.


\bibitem[int(2022)]%
        {intellicode}
 \bibinfo{year}{2022}\natexlab{}.
\newblock \bibinfo{title}{{intellicode}}.
\newblock
\newblock
\newblock
\shownote{\url{https://visualstudio.microsoft.com/services/intellicode}}.


\bibitem[ft2(2023)]%
        {ft2rawebsite}
 \bibinfo{year}{2023}\natexlab{}.
\newblock \bibinfo{title}{{ft2ra website}}.
\newblock
\newblock
\newblock
\shownote{\url{https://sites.google.com/view/ft2ra/home}}.


\bibitem[sta(2023)]%
        {stanfordcs229}
 \bibinfo{year}{2023}\natexlab{}.
\newblock \bibinfo{title}{{Stanford University CS229: Machine Learning}}.
\newblock \bibinfo{howpublished}{\url{https://cs229.stanford.edu/}}.
\newblock
\newblock
\shownote{Accessed: 2023-12-10}.


\bibitem[Ahmed et~al\mbox{.}(2024)]%
        {ahmed2024automatic}
\bibfield{author}{\bibinfo{person}{Toufique Ahmed}, \bibinfo{person}{Kunal~Suresh Pai}, \bibinfo{person}{Premkumar Devanbu}, {and} \bibinfo{person}{Earl~T Barr}.} \bibinfo{year}{2024}\natexlab{}.
\newblock \showarticletitle{Automatic semantic augmentation of language model prompts (for code summarization)}. In \bibinfo{booktitle}{\emph{2024 IEEE/ACM 46th International Conference on Software Engineering (ICSE)}}. IEEE Computer Society, \bibinfo{pages}{1004--1004}.
\newblock


\bibitem[Allamanis and Sutton(2013)]%
        {allamanis2013mining}
\bibfield{author}{\bibinfo{person}{Miltiadis Allamanis} {and} \bibinfo{person}{Charles Sutton}.} \bibinfo{year}{2013}\natexlab{}.
\newblock \showarticletitle{Mining Source Code Repositories at Massive Scale using Language Modeling}. In \bibinfo{booktitle}{\emph{2013 10th Working Conference on Mining Software Repositories (MSR)}}. IEEE, \bibinfo{pages}{207--216}.
\newblock


\bibitem[Alon et~al\mbox{.}(2022)]%
        {alon2022neuro}
\bibfield{author}{\bibinfo{person}{Uri Alon}, \bibinfo{person}{Frank Xu}, \bibinfo{person}{Junxian He}, \bibinfo{person}{Sudipta Sengupta}, \bibinfo{person}{Dan Roth}, {and} \bibinfo{person}{Graham Neubig}.} \bibinfo{year}{2022}\natexlab{}.
\newblock \showarticletitle{Neuro-symbolic language modeling with automaton-augmented retrieval}. In \bibinfo{booktitle}{\emph{International Conference on Machine Learning}}. PMLR, \bibinfo{pages}{468--485}.
\newblock


\bibitem[Borgeaud et~al\mbox{.}(2022)]%
        {borgeaud2022improving}
\bibfield{author}{\bibinfo{person}{Sebastian Borgeaud}, \bibinfo{person}{Arthur Mensch}, \bibinfo{person}{Jordan Hoffmann}, \bibinfo{person}{Trevor Cai}, \bibinfo{person}{Eliza Rutherford}, \bibinfo{person}{Katie Millican}, \bibinfo{person}{George~Bm Van Den~Driessche}, \bibinfo{person}{Jean-Baptiste Lespiau}, \bibinfo{person}{Bogdan Damoc}, \bibinfo{person}{Aidan Clark}, {et~al\mbox{.}}} \bibinfo{year}{2022}\natexlab{}.
\newblock \showarticletitle{Improving language models by retrieving from trillions of tokens}. In \bibinfo{booktitle}{\emph{International conference on machine learning}}. PMLR, \bibinfo{pages}{2206--2240}.
\newblock


\bibitem[Chen et~al\mbox{.}(2022)]%
        {chen2022decoupling}
\bibfield{author}{\bibinfo{person}{Xiang Chen}, \bibinfo{person}{Lei Li}, \bibinfo{person}{Ningyu Zhang}, \bibinfo{person}{Xiaozhuan Liang}, \bibinfo{person}{Shumin Deng}, \bibinfo{person}{Chuanqi Tan}, \bibinfo{person}{Fei Huang}, \bibinfo{person}{Luo Si}, {and} \bibinfo{person}{Huajun Chen}.} \bibinfo{year}{2022}\natexlab{}.
\newblock \showarticletitle{Decoupling knowledge from memorization: Retrieval-augmented prompt learning}.
\newblock \bibinfo{journal}{\emph{Advances in Neural Information Processing Systems}}  \bibinfo{volume}{35} (\bibinfo{year}{2022}), \bibinfo{pages}{23908--23922}.
\newblock


\bibitem[De~Jong et~al\mbox{.}(2021)]%
        {de2021mention}
\bibfield{author}{\bibinfo{person}{Michiel De~Jong}, \bibinfo{person}{Yury Zemlyanskiy}, \bibinfo{person}{Nicholas FitzGerald}, \bibinfo{person}{Fei Sha}, {and} \bibinfo{person}{William Cohen}.} \bibinfo{year}{2021}\natexlab{}.
\newblock \showarticletitle{Mention memory: incorporating textual knowledge into transformers through entity mention attention}.
\newblock \bibinfo{journal}{\emph{arXiv preprint arXiv:2110.06176}} (\bibinfo{year}{2021}).
\newblock


\bibitem[Drozdov et~al\mbox{.}(2022)]%
        {drozdov2022you}
\bibfield{author}{\bibinfo{person}{Andrew Drozdov}, \bibinfo{person}{Shufan Wang}, \bibinfo{person}{Razieh Rahimi}, \bibinfo{person}{Andrew McCallum}, \bibinfo{person}{Hamed Zamani}, {and} \bibinfo{person}{Mohit Iyyer}.} \bibinfo{year}{2022}\natexlab{}.
\newblock \showarticletitle{You can't pick your neighbors, or can you? When and how to rely on retrieval in the $ k $ NN-LM}.
\newblock \bibinfo{journal}{\emph{arXiv preprint arXiv:2210.15859}} (\bibinfo{year}{2022}).
\newblock


\bibitem[Feng et~al\mbox{.}(2020)]%
        {feng2020codebert}
\bibfield{author}{\bibinfo{person}{Zhangyin Feng}, \bibinfo{person}{Daya Guo}, \bibinfo{person}{Duyu Tang}, \bibinfo{person}{Nan Duan}, \bibinfo{person}{Xiaocheng Feng}, \bibinfo{person}{Ming Gong}, \bibinfo{person}{Linjun Shou}, \bibinfo{person}{Bing Qin}, \bibinfo{person}{Ting Liu}, \bibinfo{person}{Daxin Jiang}, {et~al\mbox{.}}} \bibinfo{year}{2020}\natexlab{}.
\newblock \showarticletitle{Codebert: A pre-trained model for programming and natural languages}.
\newblock \bibinfo{journal}{\emph{arXiv preprint arXiv:2002.08155}} (\bibinfo{year}{2020}).
\newblock


\bibitem[Guo et~al\mbox{.}(2022)]%
        {guo2022unixcoder}
\bibfield{author}{\bibinfo{person}{Daya Guo}, \bibinfo{person}{Shuai Lu}, \bibinfo{person}{Nan Duan}, \bibinfo{person}{Yanlin Wang}, \bibinfo{person}{Ming Zhou}, {and} \bibinfo{person}{Jian Yin}.} \bibinfo{year}{2022}\natexlab{}.
\newblock \showarticletitle{Unixcoder: Unified cross-modal pre-training for code representation}.
\newblock \bibinfo{journal}{\emph{arXiv preprint arXiv:2203.03850}} (\bibinfo{year}{2022}).
\newblock


\bibitem[Guu et~al\mbox{.}(2020)]%
        {guu2020retrieval}
\bibfield{author}{\bibinfo{person}{Kelvin Guu}, \bibinfo{person}{Kenton Lee}, \bibinfo{person}{Zora Tung}, \bibinfo{person}{Panupong Pasupat}, {and} \bibinfo{person}{Mingwei Chang}.} \bibinfo{year}{2020}\natexlab{}.
\newblock \showarticletitle{Retrieval augmented language model pre-training}. In \bibinfo{booktitle}{\emph{International conference on machine learning}}. PMLR, \bibinfo{pages}{3929--3938}.
\newblock


\bibitem[He et~al\mbox{.}(2021)]%
        {he2021efficient}
\bibfield{author}{\bibinfo{person}{Junxian He}, \bibinfo{person}{Graham Neubig}, {and} \bibinfo{person}{Taylor Berg-Kirkpatrick}.} \bibinfo{year}{2021}\natexlab{}.
\newblock \showarticletitle{Efficient nearest neighbor language models}.
\newblock \bibinfo{journal}{\emph{arXiv preprint arXiv:2109.04212}} (\bibinfo{year}{2021}).
\newblock


\bibitem[Hindle et~al\mbox{.}(2016)]%
        {hindle2016naturalness}
\bibfield{author}{\bibinfo{person}{Abram Hindle}, \bibinfo{person}{Earl~T Barr}, \bibinfo{person}{Mark Gabel}, \bibinfo{person}{Zhendong Su}, {and} \bibinfo{person}{Premkumar Devanbu}.} \bibinfo{year}{2016}\natexlab{}.
\newblock \showarticletitle{On the naturalness of software}.
\newblock \bibinfo{journal}{\emph{Commun. ACM}} \bibinfo{volume}{59}, \bibinfo{number}{5} (\bibinfo{year}{2016}), \bibinfo{pages}{122--131}.
\newblock


\bibitem[Hofst{\"a}tter et~al\mbox{.}(2023)]%
        {hofstatter2023fid}
\bibfield{author}{\bibinfo{person}{Sebastian Hofst{\"a}tter}, \bibinfo{person}{Jiecao Chen}, \bibinfo{person}{Karthik Raman}, {and} \bibinfo{person}{Hamed Zamani}.} \bibinfo{year}{2023}\natexlab{}.
\newblock \showarticletitle{Fid-light: Efficient and effective retrieval-augmented text generation}. In \bibinfo{booktitle}{\emph{Proceedings of the 46th International ACM SIGIR Conference on Research and Development in Information Retrieval}}. \bibinfo{pages}{1437--1447}.
\newblock


\bibitem[Hu et~al\mbox{.}(2021)]%
        {hu2021lora}
\bibfield{author}{\bibinfo{person}{Edward~J Hu}, \bibinfo{person}{Yelong Shen}, \bibinfo{person}{Phillip Wallis}, \bibinfo{person}{Zeyuan Allen-Zhu}, \bibinfo{person}{Yuanzhi Li}, \bibinfo{person}{Shean Wang}, \bibinfo{person}{Lu Wang}, {and} \bibinfo{person}{Weizhu Chen}.} \bibinfo{year}{2021}\natexlab{}.
\newblock \showarticletitle{Lora: Low-rank adaptation of large language models}.
\newblock \bibinfo{journal}{\emph{arXiv preprint arXiv:2106.09685}} (\bibinfo{year}{2021}).
\newblock


\bibitem[Husain et~al\mbox{.}(2019)]%
        {husain2019codesearchnet}
\bibfield{author}{\bibinfo{person}{Hamel Husain}, \bibinfo{person}{Ho-Hsiang Wu}, \bibinfo{person}{Tiferet Gazit}, \bibinfo{person}{Miltiadis Allamanis}, {and} \bibinfo{person}{Marc Brockschmidt}.} \bibinfo{year}{2019}\natexlab{}.
\newblock \showarticletitle{Codesearchnet challenge: Evaluating the state of semantic code search}.
\newblock \bibinfo{journal}{\emph{arXiv preprint arXiv:1909.09436}} (\bibinfo{year}{2019}).
\newblock


\bibitem[Izacard et~al\mbox{.}(2022)]%
        {izacard2022few}
\bibfield{author}{\bibinfo{person}{Gautier Izacard}, \bibinfo{person}{Patrick Lewis}, \bibinfo{person}{Maria Lomeli}, \bibinfo{person}{Lucas Hosseini}, \bibinfo{person}{Fabio Petroni}, \bibinfo{person}{Timo Schick}, \bibinfo{person}{Jane Dwivedi-Yu}, \bibinfo{person}{Armand Joulin}, \bibinfo{person}{Sebastian Riedel}, {and} \bibinfo{person}{Edouard Grave}.} \bibinfo{year}{2022}\natexlab{}.
\newblock \showarticletitle{Few-shot learning with retrieval augmented language models}.
\newblock \bibinfo{journal}{\emph{arXiv preprint arXiv:2208.03299}} (\bibinfo{year}{2022}).
\newblock


\bibitem[Jiang et~al\mbox{.}(2023)]%
        {jiang2023active}
\bibfield{author}{\bibinfo{person}{Zhengbao Jiang}, \bibinfo{person}{Frank~F Xu}, \bibinfo{person}{Luyu Gao}, \bibinfo{person}{Zhiqing Sun}, \bibinfo{person}{Qian Liu}, \bibinfo{person}{Jane Dwivedi-Yu}, \bibinfo{person}{Yiming Yang}, \bibinfo{person}{Jamie Callan}, {and} \bibinfo{person}{Graham Neubig}.} \bibinfo{year}{2023}\natexlab{}.
\newblock \showarticletitle{Active retrieval augmented generation}.
\newblock \bibinfo{journal}{\emph{arXiv preprint arXiv:2305.06983}} (\bibinfo{year}{2023}).
\newblock


\bibitem[Joshi et~al\mbox{.}(2023)]%
        {joshi2023repair}
\bibfield{author}{\bibinfo{person}{Harshit Joshi}, \bibinfo{person}{Jos{\'e}~Cambronero Sanchez}, \bibinfo{person}{Sumit Gulwani}, \bibinfo{person}{Vu Le}, \bibinfo{person}{Gust Verbruggen}, {and} \bibinfo{person}{Ivan Radi{\v{c}}ek}.} \bibinfo{year}{2023}\natexlab{}.
\newblock \showarticletitle{Repair is nearly generation: Multilingual program repair with llms}. In \bibinfo{booktitle}{\emph{Proceedings of the AAAI Conference on Artificial Intelligence}}, Vol.~\bibinfo{volume}{37}. \bibinfo{pages}{5131--5140}.
\newblock


\bibitem[Karpukhin et~al\mbox{.}(2020)]%
        {karpukhin2020dense}
\bibfield{author}{\bibinfo{person}{Vladimir Karpukhin}, \bibinfo{person}{Barlas O{\u{g}}uz}, \bibinfo{person}{Sewon Min}, \bibinfo{person}{Patrick Lewis}, \bibinfo{person}{Ledell Wu}, \bibinfo{person}{Sergey Edunov}, \bibinfo{person}{Danqi Chen}, {and} \bibinfo{person}{Wen-tau Yih}.} \bibinfo{year}{2020}\natexlab{}.
\newblock \showarticletitle{Dense passage retrieval for open-domain question answering}.
\newblock \bibinfo{journal}{\emph{arXiv preprint arXiv:2004.04906}} (\bibinfo{year}{2020}).
\newblock


\bibitem[Khandelwal et~al\mbox{.}(2020)]%
        {Khandelwal2020Generalization}
\bibfield{author}{\bibinfo{person}{Urvashi Khandelwal}, \bibinfo{person}{Omer Levy}, \bibinfo{person}{Dan Jurafsky}, \bibinfo{person}{Luke Zettlemoyer}, {and} \bibinfo{person}{Mike Lewis}.} \bibinfo{year}{2020}\natexlab{}.
\newblock \showarticletitle{Generalization through Memorization: Nearest Neighbor Language Models}. In \bibinfo{booktitle}{\emph{International Conference on Learning Representations}}.
\newblock
\urldef\tempurl%
\url{https://openreview.net/forum?id=HklBjCEKvH}
\showURL{%
\tempurl}


\bibitem[Kim et~al\mbox{.}(2021)]%
        {kim2021code}
\bibfield{author}{\bibinfo{person}{Seohyun Kim}, \bibinfo{person}{Jinman Zhao}, \bibinfo{person}{Yuchi Tian}, {and} \bibinfo{person}{Satish Chandra}.} \bibinfo{year}{2021}\natexlab{}.
\newblock \showarticletitle{Code prediction by feeding trees to transformers}. In \bibinfo{booktitle}{\emph{2021 IEEE/ACM 43rd International Conference on Software Engineering (ICSE)}}. IEEE, \bibinfo{pages}{150--162}.
\newblock


\bibitem[Lester et~al\mbox{.}(2021)]%
        {lester2021power}
\bibfield{author}{\bibinfo{person}{Brian Lester}, \bibinfo{person}{Rami Al-Rfou}, {and} \bibinfo{person}{Noah Constant}.} \bibinfo{year}{2021}\natexlab{}.
\newblock \showarticletitle{The power of scale for parameter-efficient prompt tuning}.
\newblock \bibinfo{journal}{\emph{arXiv preprint arXiv:2104.08691}} (\bibinfo{year}{2021}).
\newblock


\bibitem[Lewis et~al\mbox{.}(2020)]%
        {lewis2020retrieval}
\bibfield{author}{\bibinfo{person}{Patrick Lewis}, \bibinfo{person}{Ethan Perez}, \bibinfo{person}{Aleksandra Piktus}, \bibinfo{person}{Fabio Petroni}, \bibinfo{person}{Vladimir Karpukhin}, \bibinfo{person}{Naman Goyal}, \bibinfo{person}{Heinrich K{\"u}ttler}, \bibinfo{person}{Mike Lewis}, \bibinfo{person}{Wen-tau Yih}, \bibinfo{person}{Tim Rockt{\"a}schel}, {et~al\mbox{.}}} \bibinfo{year}{2020}\natexlab{}.
\newblock \showarticletitle{Retrieval-augmented generation for knowledge-intensive nlp tasks}.
\newblock \bibinfo{journal}{\emph{Advances in Neural Information Processing Systems}}  \bibinfo{volume}{33} (\bibinfo{year}{2020}), \bibinfo{pages}{9459--9474}.
\newblock


\bibitem[Li et~al\mbox{.}(2017)]%
        {li2017code}
\bibfield{author}{\bibinfo{person}{Jian Li}, \bibinfo{person}{Yue Wang}, \bibinfo{person}{Michael~R Lyu}, {and} \bibinfo{person}{Irwin King}.} \bibinfo{year}{2017}\natexlab{}.
\newblock \showarticletitle{Code completion with neural attention and pointer networks}.
\newblock \bibinfo{journal}{\emph{arXiv preprint arXiv:1711.09573}} (\bibinfo{year}{2017}).
\newblock


\bibitem[Li et~al\mbox{.}(2023)]%
        {li2023starcoder}
\bibfield{author}{\bibinfo{person}{Raymond Li}, \bibinfo{person}{Loubna~Ben Allal}, \bibinfo{person}{Yangtian Zi}, \bibinfo{person}{Niklas Muennighoff}, \bibinfo{person}{Denis Kocetkov}, \bibinfo{person}{Chenghao Mou}, \bibinfo{person}{Marc Marone}, \bibinfo{person}{Christopher Akiki}, \bibinfo{person}{Jia Li}, \bibinfo{person}{Jenny Chim}, {et~al\mbox{.}}} \bibinfo{year}{2023}\natexlab{}.
\newblock \showarticletitle{StarCoder: may the source be with you!}
\newblock \bibinfo{journal}{\emph{arXiv preprint arXiv:2305.06161}} (\bibinfo{year}{2023}).
\newblock


\bibitem[Liu et~al\mbox{.}(2016)]%
        {liu2016neural}
\bibfield{author}{\bibinfo{person}{Chang Liu}, \bibinfo{person}{Xin Wang}, \bibinfo{person}{Richard Shin}, \bibinfo{person}{Joseph~E Gonzalez}, {and} \bibinfo{person}{Dawn Song}.} \bibinfo{year}{2016}\natexlab{}.
\newblock \showarticletitle{Neural code completion}.
\newblock  (\bibinfo{year}{2016}).
\newblock


\bibitem[Liu et~al\mbox{.}(2020a)]%
        {liu2020retrieval}
\bibfield{author}{\bibinfo{person}{Shangqing Liu}, \bibinfo{person}{Yu Chen}, \bibinfo{person}{Xiaofei Xie}, \bibinfo{person}{Jingkai Siow}, {and} \bibinfo{person}{Yang Liu}.} \bibinfo{year}{2020}\natexlab{a}.
\newblock \showarticletitle{Retrieval-augmented generation for code summarization via hybrid gnn}.
\newblock \bibinfo{journal}{\emph{arXiv preprint arXiv:2006.05405}} (\bibinfo{year}{2020}).
\newblock


\bibitem[Liu et~al\mbox{.}(2020b)]%
        {liu2020atom}
\bibfield{author}{\bibinfo{person}{Shangqing Liu}, \bibinfo{person}{Cuiyun Gao}, \bibinfo{person}{Sen Chen}, \bibinfo{person}{Lun~Yiu Nie}, {and} \bibinfo{person}{Yang Liu}.} \bibinfo{year}{2020}\natexlab{b}.
\newblock \showarticletitle{Atom: Commit message generation based on abstract syntax tree and hybrid ranking}.
\newblock \bibinfo{journal}{\emph{IEEE Transactions on Software Engineering}} \bibinfo{volume}{48}, \bibinfo{number}{5} (\bibinfo{year}{2020}), \bibinfo{pages}{1800--1817}.
\newblock


\bibitem[Liu et~al\mbox{.}(2022)]%
        {liu2022commitbart}
\bibfield{author}{\bibinfo{person}{Shangqing Liu}, \bibinfo{person}{Yanzhou Li}, \bibinfo{person}{Xiaofei Xie}, {and} \bibinfo{person}{Yang Liu}.} \bibinfo{year}{2022}\natexlab{}.
\newblock \showarticletitle{Commitbart: A large pre-trained model for github commits}.
\newblock \bibinfo{journal}{\emph{arXiv preprint arXiv:2208.08100}} (\bibinfo{year}{2022}).
\newblock


\bibitem[Liu et~al\mbox{.}(2023a)]%
        {liu2023contrabert}
\bibfield{author}{\bibinfo{person}{Shangqing Liu}, \bibinfo{person}{Bozhi Wu}, \bibinfo{person}{Xiaofei Xie}, \bibinfo{person}{Guozhu Meng}, {and} \bibinfo{person}{Yang Liu}.} \bibinfo{year}{2023}\natexlab{a}.
\newblock \showarticletitle{Contrabert: Enhancing code pre-trained models via contrastive learning}. In \bibinfo{booktitle}{\emph{2023 IEEE/ACM 45th International Conference on Software Engineering (ICSE)}}. IEEE, \bibinfo{pages}{2476--2487}.
\newblock


\bibitem[Liu et~al\mbox{.}(2023b)]%
        {liu2023gpt}
\bibfield{author}{\bibinfo{person}{Xiao Liu}, \bibinfo{person}{Yanan Zheng}, \bibinfo{person}{Zhengxiao Du}, \bibinfo{person}{Ming Ding}, \bibinfo{person}{Yujie Qian}, \bibinfo{person}{Zhilin Yang}, {and} \bibinfo{person}{Jie Tang}.} \bibinfo{year}{2023}\natexlab{b}.
\newblock \showarticletitle{GPT understands, too}.
\newblock \bibinfo{journal}{\emph{AI Open}} (\bibinfo{year}{2023}).
\newblock


\bibitem[Lu et~al\mbox{.}(2022)]%
        {lu2022reacc}
\bibfield{author}{\bibinfo{person}{Shuai Lu}, \bibinfo{person}{Nan Duan}, \bibinfo{person}{Hojae Han}, \bibinfo{person}{Daya Guo}, \bibinfo{person}{Seung-won Hwang}, {and} \bibinfo{person}{Alexey Svyatkovskiy}.} \bibinfo{year}{2022}\natexlab{}.
\newblock \showarticletitle{Reacc: A retrieval-augmented code completion framework}.
\newblock \bibinfo{journal}{\emph{arXiv preprint arXiv:2203.07722}} (\bibinfo{year}{2022}).
\newblock


\bibitem[Lu et~al\mbox{.}(2021)]%
        {lu2021codexglue}
\bibfield{author}{\bibinfo{person}{Shuai Lu}, \bibinfo{person}{Daya Guo}, \bibinfo{person}{Shuo Ren}, \bibinfo{person}{Junjie Huang}, \bibinfo{person}{Alexey Svyatkovskiy}, \bibinfo{person}{Ambrosio Blanco}, \bibinfo{person}{Colin Clement}, \bibinfo{person}{Dawn Drain}, \bibinfo{person}{Daxin Jiang}, \bibinfo{person}{Duyu Tang}, {et~al\mbox{.}}} \bibinfo{year}{2021}\natexlab{}.
\newblock \showarticletitle{Codexglue: A machine learning benchmark dataset for code understanding and generation}.
\newblock \bibinfo{journal}{\emph{arXiv preprint arXiv:2102.04664}} (\bibinfo{year}{2021}).
\newblock


\bibitem[Nijkamp et~al\mbox{.}(2022)]%
        {nijkamp2022codegen}
\bibfield{author}{\bibinfo{person}{Erik Nijkamp}, \bibinfo{person}{Bo Pang}, \bibinfo{person}{Hiroaki Hayashi}, \bibinfo{person}{Lifu Tu}, \bibinfo{person}{Huan Wang}, \bibinfo{person}{Yingbo Zhou}, \bibinfo{person}{Silvio Savarese}, {and} \bibinfo{person}{Caiming Xiong}.} \bibinfo{year}{2022}\natexlab{}.
\newblock \showarticletitle{Codegen: An open large language model for code with multi-turn program synthesis}.
\newblock \bibinfo{journal}{\emph{arXiv preprint arXiv:2203.13474}} (\bibinfo{year}{2022}).
\newblock


\bibitem[OpenAI(2023)]%
        {ChatGPTblog}
\bibfield{author}{\bibinfo{person}{OpenAI}.} \bibinfo{year}{2023}\natexlab{}.
\newblock \bibinfo{title}{ChatGPTblog}.
\newblock \bibinfo{howpublished}{\url{https://openai.com/blog/chatgpt}}.
\newblock


\bibitem[Parvez et~al\mbox{.}(2021)]%
        {parvez2021retrieval}
\bibfield{author}{\bibinfo{person}{Md~Rizwan Parvez}, \bibinfo{person}{Wasi~Uddin Ahmad}, \bibinfo{person}{Saikat Chakraborty}, \bibinfo{person}{Baishakhi Ray}, {and} \bibinfo{person}{Kai-Wei Chang}.} \bibinfo{year}{2021}\natexlab{}.
\newblock \showarticletitle{Retrieval augmented code generation and summarization}.
\newblock \bibinfo{journal}{\emph{arXiv preprint arXiv:2108.11601}} (\bibinfo{year}{2021}).
\newblock


\bibitem[Radford et~al\mbox{.}(2019)]%
        {radford2019language}
\bibfield{author}{\bibinfo{person}{Alec Radford}, \bibinfo{person}{Jeffrey Wu}, \bibinfo{person}{Rewon Child}, \bibinfo{person}{David Luan}, \bibinfo{person}{Dario Amodei}, \bibinfo{person}{Ilya Sutskever}, {et~al\mbox{.}}} \bibinfo{year}{2019}\natexlab{}.
\newblock \showarticletitle{Language models are unsupervised multitask learners}.
\newblock \bibinfo{journal}{\emph{OpenAI blog}} \bibinfo{volume}{1}, \bibinfo{number}{8} (\bibinfo{year}{2019}), \bibinfo{pages}{9}.
\newblock


\bibitem[Ram et~al\mbox{.}(2023)]%
        {ram2023context}
\bibfield{author}{\bibinfo{person}{Ori Ram}, \bibinfo{person}{Yoav Levine}, \bibinfo{person}{Itay Dalmedigos}, \bibinfo{person}{Dor Muhlgay}, \bibinfo{person}{Amnon Shashua}, \bibinfo{person}{Kevin Leyton-Brown}, {and} \bibinfo{person}{Yoav Shoham}.} \bibinfo{year}{2023}\natexlab{}.
\newblock \showarticletitle{In-context retrieval-augmented language models}.
\newblock \bibinfo{journal}{\emph{arXiv preprint arXiv:2302.00083}} (\bibinfo{year}{2023}).
\newblock


\bibitem[Raychev et~al\mbox{.}(2016)]%
        {raychev2016probabilistic}
\bibfield{author}{\bibinfo{person}{Veselin Raychev}, \bibinfo{person}{Pavol Bielik}, {and} \bibinfo{person}{Martin Vechev}.} \bibinfo{year}{2016}\natexlab{}.
\newblock \showarticletitle{Probabilistic Model for Code with Decision Trees}.
\newblock \bibinfo{journal}{\emph{ACM SIGPLAN Notices}} (\bibinfo{year}{2016}), \bibinfo{pages}{731--747}.
\newblock


\bibitem[Robertson et~al\mbox{.}(2009)]%
        {robertson2009probabilistic}
\bibfield{author}{\bibinfo{person}{Stephen Robertson}, \bibinfo{person}{Hugo Zaragoza}, {et~al\mbox{.}}} \bibinfo{year}{2009}\natexlab{}.
\newblock \showarticletitle{The probabilistic relevance framework: BM25 and beyond}.
\newblock \bibinfo{journal}{\emph{Foundations and Trends{\textregistered} in Information Retrieval}} \bibinfo{volume}{3}, \bibinfo{number}{4} (\bibinfo{year}{2009}), \bibinfo{pages}{333--389}.
\newblock


\bibitem[Shi et~al\mbox{.}(2023)]%
        {shi2023replug}
\bibfield{author}{\bibinfo{person}{Weijia Shi}, \bibinfo{person}{Sewon Min}, \bibinfo{person}{Michihiro Yasunaga}, \bibinfo{person}{Minjoon Seo}, \bibinfo{person}{Rich James}, \bibinfo{person}{Mike Lewis}, \bibinfo{person}{Luke Zettlemoyer}, {and} \bibinfo{person}{Wen-tau Yih}.} \bibinfo{year}{2023}\natexlab{}.
\newblock \showarticletitle{Replug: Retrieval-augmented black-box language models}.
\newblock \bibinfo{journal}{\emph{arXiv preprint arXiv:2301.12652}} (\bibinfo{year}{2023}).
\newblock


\bibitem[Shrivastava et~al\mbox{.}(2023)]%
        {shrivastava2023repository}
\bibfield{author}{\bibinfo{person}{Disha Shrivastava}, \bibinfo{person}{Hugo Larochelle}, {and} \bibinfo{person}{Daniel Tarlow}.} \bibinfo{year}{2023}\natexlab{}.
\newblock \showarticletitle{Repository-level prompt generation for large language models of code}. In \bibinfo{booktitle}{\emph{International Conference on Machine Learning}}. PMLR, \bibinfo{pages}{31693--31715}.
\newblock


\bibitem[Tang et~al\mbox{.}(2023)]%
        {tang2023domain}
\bibfield{author}{\bibinfo{person}{Ze Tang}, \bibinfo{person}{Jidong Ge}, \bibinfo{person}{Shangqing Liu}, \bibinfo{person}{Tingwei Zhu}, \bibinfo{person}{Tongtong Xu}, \bibinfo{person}{Liguo Huang}, {and} \bibinfo{person}{Bin Luo}.} \bibinfo{year}{2023}\natexlab{}.
\newblock \showarticletitle{Domain Adaptive Code Completion via Language Models and Decoupled Domain Databases}.
\newblock \bibinfo{journal}{\emph{arXiv preprint arXiv:2308.09313}} (\bibinfo{year}{2023}).
\newblock


\bibitem[Wang et~al\mbox{.}(2021)]%
        {wang2021codet5}
\bibfield{author}{\bibinfo{person}{Yue Wang}, \bibinfo{person}{Weishi Wang}, \bibinfo{person}{Shafiq Joty}, {and} \bibinfo{person}{Steven~CH Hoi}.} \bibinfo{year}{2021}\natexlab{}.
\newblock \showarticletitle{Codet5: Identifier-aware unified pre-trained encoder-decoder models for code understanding and generation}.
\newblock \bibinfo{journal}{\emph{arXiv preprint arXiv:2109.00859}} (\bibinfo{year}{2021}).
\newblock


\bibitem[Wang et~al\mbox{.}(2023)]%
        {wang2023learning}
\bibfield{author}{\bibinfo{person}{Zhiruo Wang}, \bibinfo{person}{Jun Araki}, \bibinfo{person}{Zhengbao Jiang}, \bibinfo{person}{Md~Rizwan Parvez}, {and} \bibinfo{person}{Graham Neubig}.} \bibinfo{year}{2023}\natexlab{}.
\newblock \showarticletitle{Learning to Filter Context for Retrieval-Augmented Generation}.
\newblock \bibinfo{journal}{\emph{arXiv preprint arXiv:2311.08377}} (\bibinfo{year}{2023}).
\newblock


\bibitem[Wikipedia(2023a)]%
        {WikipediaEmpiricalProb}
\bibfield{author}{\bibinfo{person}{Wikipedia}.} \bibinfo{year}{2023}\natexlab{a}.
\newblock \bibinfo{title}{Empirical Probability}.
\newblock \bibinfo{howpublished}{\url{https://en.wikipedia.org/wiki/Empirical_probability}}.
\newblock


\bibitem[Wikipedia(2023b)]%
        {WikipediaFrequency}
\bibfield{author}{\bibinfo{person}{Wikipedia}.} \bibinfo{year}{2023}\natexlab{b}.
\newblock \bibinfo{title}{Frequency (statistics)}.
\newblock \bibinfo{howpublished}{\url{https://en.wikipedia.org/wiki/Frequency_(statistics)}}.
\newblock


\bibitem[Xu et~al\mbox{.}(2023)]%
        {xu2023nearest}
\bibfield{author}{\bibinfo{person}{Frank~F Xu}, \bibinfo{person}{Uri Alon}, {and} \bibinfo{person}{Graham Neubig}.} \bibinfo{year}{2023}\natexlab{}.
\newblock \showarticletitle{Why do Nearest Neighbor Language Models Work?}
\newblock \bibinfo{journal}{\emph{arXiv preprint arXiv:2301.02828}} (\bibinfo{year}{2023}).
\newblock


\bibitem[Yang et~al\mbox{.}(2023)]%
        {yang2023leandojo}
\bibfield{author}{\bibinfo{person}{Kaiyu Yang}, \bibinfo{person}{Aidan~M Swope}, \bibinfo{person}{Alex Gu}, \bibinfo{person}{Rahul Chalamala}, \bibinfo{person}{Peiyang Song}, \bibinfo{person}{Shixing Yu}, \bibinfo{person}{Saad Godil}, \bibinfo{person}{Ryan Prenger}, {and} \bibinfo{person}{Anima Anandkumar}.} \bibinfo{year}{2023}\natexlab{}.
\newblock \showarticletitle{Leandojo: Theorem proving with retrieval-augmented language models}.
\newblock \bibinfo{journal}{\emph{arXiv preprint arXiv:2306.15626}} (\bibinfo{year}{2023}).
\newblock


\bibitem[Zhang et~al\mbox{.}(2023)]%
        {zhang2023repocoder}
\bibfield{author}{\bibinfo{person}{Fengji Zhang}, \bibinfo{person}{Bei Chen}, \bibinfo{person}{Yue Zhang}, \bibinfo{person}{Jin Liu}, \bibinfo{person}{Daoguang Zan}, \bibinfo{person}{Yi Mao}, \bibinfo{person}{Jian-Guang Lou}, {and} \bibinfo{person}{Weizhu Chen}.} \bibinfo{year}{2023}\natexlab{}.
\newblock \showarticletitle{Repocoder: Repository-level code completion through iterative retrieval and generation}.
\newblock \bibinfo{journal}{\emph{arXiv preprint arXiv:2303.12570}} (\bibinfo{year}{2023}).
\newblock


\end{thebibliography}
